\documentclass[prb,dvipdfm,floatfix,12pt,letterpaper]{revtex4}

\usepackage{graphicx,amssymb,amsfonts,amsmath,color,subfigure}
\usepackage[colorlinks=true,linkcolor=blue,citecolor=red]{hyperref}

\begin{document}

\title{The role of viscoelastic contrast in orientation selection of block
copolymer lamellar phases under oscillatory shear}

\author{Chi-Deuk Yoo} \author{Jorge Vi\~nals}

\affiliation{School of Physics and Astronomy, and Minnesota Supercomputing
Institute, University of Minnesota, 116 Church Street S.E., Minneapolis, MN
55455}

\date{\today}

\begin{abstract} 

The mesoscale rheology of a lamellar phase of a block copolymer is modeled as a
structured fluid of uniaxial symmetry. The model predicts a viscoelastic
response that depends on the angle between the the local lamellar planes and
velocity gradients. We focus on the stability under oscillatory shear of a two
layer configuration comprising a parallel and a perpendicularly oriented
domain, so that the two layers have a different viscoelastic modulus
$G^{*}(\omega)$.  A long wave, low Reynolds number expansion is introduced to
analytically obtain the region of stability. When the response of the two
layers is purely viscous, we recover earlier results according to which the
interface is unstable for non zero Reynolds number flows when the thinner layer
is more viscous. On the other hand, when viscoelasticity is included, we find
that the interface can become unstable even for zero Reynolds number. The
interfacial instability is argued to dynamically favor perpendicular relative
to parallel orientation, and hence we suggest that the perpendicular
orientation would be selected in a multi domain configuration in the range of
frequency $\omega$ in which viscoelastic contrast among orientations is
appreciable.  

\end{abstract}

\maketitle

\section{Introduction}

A stability analysis of two superposed fluid layers to oscillatory shear is
given when the frequency dependent viscoelastic modulus of each layer is
different. At vanishing Reynolds number, we find that an initially planar
interface can be rendered linearly unstable to long wavelength perturbations by
viscoelastic contrast. Our analysis is motivated by rheology and alignment
studies of structured phases, mostly in soft and biological matter, in which
local viscoelastic response couples to a broken symmetry variable of the phase
(e.g., orientation). The specific case considered is related to alignment
studies of lamellar phases in bulk block copolymer melts. There by reason of
symmetry, the complex viscoelastic modulus $G^{*}(\omega)$ depends on the
relative orientation between local lamellar normal and velocity gradient. We
argue that given the dependence of $G^{*}(\omega)$ on orientation, the
hydrodynamic instability discussed here would dynamically favor domains of
perpendicular orientation over parallel in a bulk, multi domain configuration.

Block copolymers are currently being investigated for a wide variety of
applications in nanotechnology \cite{park03,re:black05, re:zschech07,tang08,
re:pujari12,re:xie13} due to their ability to self-assemble into ordered
nanophases of different symmetry (lamellar, cylindrical, spherical, close
packed spherical, or bi-continuous phases such as gyroid).
\cite{re:yamada04,re:guo08,re:lee10} Equilibrium morphologies and
characteristic length scales can be easily controlled by chemical manipulation
of the polymer blocks. In general, however, quenched block copolymer samples
are macroscopically disordered in that they contain a large number of domains
of different orientation, all degenerate by symmetry.  Given that disordered or
defected samples are not generally suitable for applications
\cite{re:harrison00b,re:kim03,re:kramer05,re:bosworth07}, a number of
strategies have been put forward to accelerate annealing of defects, or to post
process polycrystalline samples in order to increase the characteristic size of
ordered domains. Applied shears have been particularly effective in aligning
thin films of in-plane cylinder \cite{re:angelescu04}, sphere forming
\cite{re:angelescu05}, and lamellar \cite{re:pujari12} phases.  The same
strategy has been used on bulk samples
\cite{re:zhang95b,re:patel95,chen98,wu05}, although there is no explanation yet
as to the selected orientation relative to the shear as a function of material
parameters, and amplitude and frequency of the shear \cite{wu05}. This lack of
understanding of the non equilibrium processes involved in the alignment of
bulk samples has led to alternative strategies including, for example, solvent
annealing and applied electric fields \cite{re:olszowka09}.


Our current understanding of shear alignment in bulk samples is well summarized
by Wu et al. \cite{wu05}.  They conducted a systematic analysis in
styrene-isoprene (SI) multiblocks (from diblocks to undecablocks).  Results for
diblocks confirm earlier experimental findings that show a sequence of
transitions from parallel orientation at low frequencies, to perpendicular, and
again to parallel as the frequency is increased.  The sequence of transitions
is qualitatively different in poly-(ethylene-propylene) - poly-(ethylethylene)
(PEP-PEE) diblocks, presumably because a marked viscoelastic contrast between
styrene and isoprene blocks which is largely absent in PEP-PEE. This issue was
already emphasized by Fredrickson \cite{re:fredrickson96} who noted that
styrene is largely unentangled, whereas isoprene is entangled, and hence a
large contrast in relaxation times of the blocks is anticipated. Further, the
experiments of Wu et al.\cite{wu05} show that this double transition is
observed in diblocks and triblocks, but surprisingly not in higher order
multiblocks, in which entanglement effects among longer chains would be
expected to be more important.

A further complicating factor in the elucidation of the orientation selected
under shear is the non terminal viscoelastic response of the block copolymer
melt in the range of frequencies in which orientation switching is observed.
\cite{koppi92, larson93, gupta95, wu05} (terminal behavior has an elastic
modulus that vanishes with frequency as $G^\prime \sim \omega^{2}$ whereas the
loss modulus reflects Newtonian viscous dissipation $G^{\prime \prime} \sim
\omega$). In fact, non terminal rheology is observed even at the lowest
frequencies at which $G^{*}(\omega)$ has been measured, a range well below that
which can be attributed to individual chain dynamics.\cite{rosedale90}
Interestingly, a dependence of the non terminal viscoelasticity on lamellar
orientation has also been found in largely oriented samples. \cite{koppi92,
larson93, gupta95, wu05}  Except for diblocks, all multiblocks in Ref.
\onlinecite{wu05} show terminal behavior as $\omega \rightarrow 0$ when aligned
along the perpendicular orientation, whereas none of them are terminal in the
parallel orientation. This observation, together with the prominent
entanglement effects described above, calls into question existing theoretical
analyses that either neglect hydrodynamics or, if they don't, they assume
Newtonian flows.  Viscoelastic response appears to be closely correlated with
orientation, and is certainly not negligible even at the lowest frequencies
probed.

We propose a new viscoelastic model of block copolymer melts that explicitly
takes into account the fact that the melt in a lamellar phase is a uniaxial (not
isotropic) fluid. As such, local viscoelastic response involves the coupling
between local orientation of the lamellae and local strain and velocity
gradients. For an incompressible system, such a symmetry requires three
viscosity coefficients and three elastic moduli \cite{re:martin72}.  We extend
here the earlier analysis of a purely viscous uniaxial fluid \cite{huang07}, 
to incorporate viscoelasticity. As shown below, a
uniaxial viscoelastic medium readily allows for differential rheology in, say,
parallel and perpendicular orientations, including the observed terminal
behavior of the perpendicular orientation versus viscoelastic response of other
orientations. 

For simplicity, our analysis does not address a macroscopically disordered
sample, but rather a configuration comprising only two perfectly oriented
domains and the boundary separating them. Of the three possible orientations of
an ordered lamellar domain relative to the imposed shear we focus here on
parallel and perpendicular only. The third independent component (the so called
transverse, in which the lamellar normal is parallel to the velocity direction),
is less stable as it is being compressed by the shear \cite{re:chen02}. Both
parallel and perpendicularly oriented domains are marginal with respect to the
flow, and as such have been the subject of most orientation selection studies to
date.

Section \ref{sec:model} describes the model equations and the fluid
configuration under consideration. We introduce a general linear viscoelastic
constitutive law for a incompressible system with uniaxial symmetry, and derive
the reference flow under oscillatory shear and a planar boundary separating two
layers of different orientation. Section \ref{sec:linear} presents our results
concerning the linear instability of the base state by considering small
perturbations of large wavelength, also in the limit of small Reynolds number.
The implications of our findings on orientation selection in a lamellar block
copolymer are discussed in Sec. \ref{sec:discussion}.

\section{Viscoelastic mesoscopic model}
\label{sec:model}

Although the description below can be generalized to the case of a multi block
(see, e.g., Ref. \onlinecite{re:xie13}), we start by introducing the two fluid
model of a diblock copolymer for monomers of type A and B. At sufficiently low
frequencies so that the individual polymer chains have relaxed to their
equilibrium conformation, the state of the copolymer can be described by a local
order parameter $\psi \sim \rho_\text{A} - \rho_{B}$, where $\rho_{A}$ and
$\rho_{B}$ are the number fractions of monomers A and B respectively. The
evolution of the order parameter has a relaxational component that is driven by
free energy reduction \cite{re:fredrickson94} If fluid flow
is allowed, there is a reversible contribution to the order parameter equation
that incorporates advection by the local flow. If $v_i$ is the $i-th$ 
component of
the local fluid velocity, the advection term for an incompressible fluid is $v_i
\partial_i \psi$, where sum over repeated indices is assumed. The flow velocity
$v_{i}$ satisfies the momentum conservation equation with the stress tensor
being the momentum density current. We now introduce a general linear
constitutive law for the stress tensor of the form, 
\begin{equation} 
\sigma_{ij}
= \int_{-\infty}^t dt^\prime \; G_{ijkl}(t-t^\prime) \Big[ \partial_k
v_l(t^\prime) + \partial_l v_k(t^\prime) \Big].  
\label{constitutive.relation}
\end{equation} 
The independent contributions to a fourth rank tensor compatible
with incompressible condition ($\delta_{ij} \partial_i v_j = 0$) and 
uniaxial symmetry ($\hat{q}_i$ is the $i-th$ component of the local unit 
normal to the lamellar planes, and assume that the constitutive law is 
invariant under $\hat{q}_{i} \rightarrow - \hat{q}_{i}$) can be written as 
\begin{equation} 
\begin{split}
G_{ijkl}(t) =& G_{1}(t) \hat{q}_i \hat{q}_j \hat{q}_k \hat{q}_l + G_{4}(t) (
\delta_{ik} \delta_{jl} + \delta_{il} \delta_{jk} ) \\ & + G_{56}(t) \Big[
\hat{q}_i ( \hat{q}_k \delta_{jl} + \hat{q}_l \delta_{kj}) + \hat{q}_j (
\hat{q}_k \delta_{il} + \hat{q}_l \delta_{ki}) \Big], \label{eq:uniaxial_fluid}
\end{split} 
\end{equation} 
where $\delta_{ij}$ is the Kronecker's delta. There are, in this case, three 
independent modulus components $G_{1}, G_{4}$ and $G_{56}$ and, as a 
consequence, the viscoelastic response depends on the local lamellar 
orientation. For example, for perfect parallel lamellae ($\hat{\bf q}
\parallel \nabla v$) $G_\parallel(t) = G_{4}(t) + G_{56}(t)$, and for perfect
perpendicular lamellae ($\hat{\bf q} \parallel \nabla \times {\bf v}$)
$G_\perp(t) = G_4(t)$. Therefore, any viscoelastic contrast between these two
orientations is contained in the component $G_{56}(t)$. Note, in particular,
that if the response of the material in the parallel configuration is terminal,
this has to be the case for a perpendicular orientation as well. The reverse,
however, is not true, fact that is consistent with experiments \cite{wu05}.

\begin{figure}[t]
\centering
\includegraphics[width=12cm]{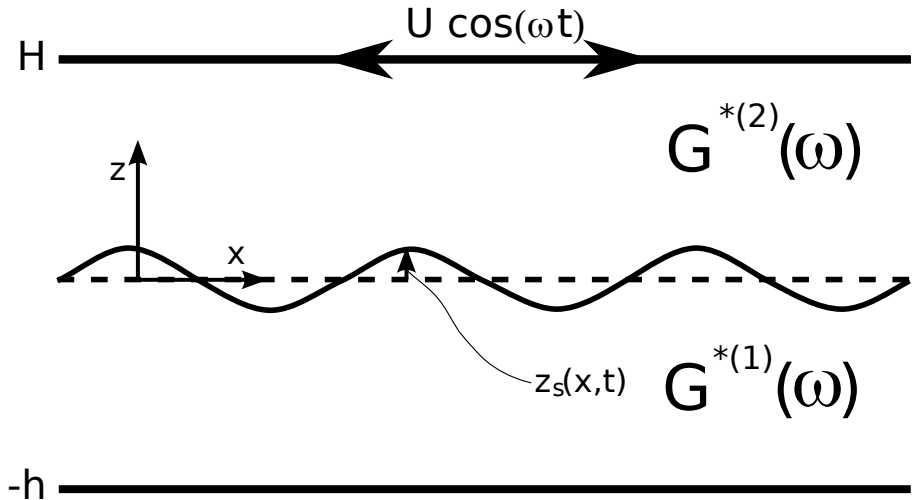}
\caption{Schematic configuration studied. As the top and bottom fluids represent
lamellar phases of parallel or perpendicular orientation, the complex modulus 
$G^{*(\alpha)}(\omega)$ of the two is different. The distortion of interface 
away from planarity is given by $z_s = z_s(x,t)$.} 
\label{fig1.schematic}
\end{figure}

We focus here on extending the results of Ref.~\onlinecite{huang07} to include 
uniaxial viscoelasticity. We consider two semi infinite layers
of a lamellar block copolymer, one on top of the other
(Fig.~\ref{fig1.schematic}), confined between two infinite and parallel planes.
The phase at the top has thickness $H$, and the lower phase $h$. The upper plane
oscillates with a velocity ${\bf U}$ of frequency $\omega$ and amplitude $U$
while the lower plane is at rest. The densities of two layers, $\rho$,
are constant and equal since
they are the same copolymer, differing only on lamellar orientation. For
simplicity, we neglect any variation in the vorticity direction ($\nabla \times
{\bf U}$),  or $\hat{y}$ in Fig.~\ref{fig1.schematic}. Hence we focus on a two
dimensional system in which the fluid velocity has components $v_i^{(\alpha)} =
(u^{(\alpha)}, 0, w^{(\alpha)})$ with $\alpha = 1, 2$ denoting fluid 1 (lower)
or 2 (upper).

When the two layers in this configuration contain parallel and
perpendicular phases, the state of the system is unaffected by the flow 
since ${\mathbf
v} \cdot \nabla \psi = 0$ for both orientations. Furthermore, if as is
customary, both phases are assumed to be isotropic fluids, then the
configuration shown is stable, the order parameter is stationary, and the flow
is a simple linear shear. If, on the other 
hand, Eq.~(\ref{eq:uniaxial_fluid}) is assumed, then the two layers have a 
different viscoelastic modulus and the configuration can be linearly unstable. 
This instability is the subject of our study below.

The response of stratified fluids to shear flows has been studied in detail.
Following pioneering work on the interfacial stability of two superposed,
incompressible and Newtonian fluids under steady and oscillatory shears by Yih
\cite{yih67, yih68}, there have been a series of additional studies involving 
viscous fluids under steady \cite{hooper83,hinch84} and oscillatory shears \cite{king99}, or
viscoelastic fluids under steady shear \cite{li69, waters87, renardy88, chen91}.
To our knowledge, however, there has been no stability analysis of stratified
viscoelastic fluids subjected to oscillatory shears. In the case of purely
viscous fluids, a stratified configuration can be unstable because of a phase
lag between the velocity fields on either side of the interface. For this to be
possible in a Newtonian fluid, the Reynolds number must be finite. This case is
unlikely to be of relevance to block copolymers as typical Reynolds numbers are
extremely small. In the viscoelastic case, on the other hand, a differential
elastic response between the two phases, even at zero Reynolds number, provides
for a mechanism for instability as detailed below.

In each fluid layer, the velocity field satisfies
\begin{equation}
\rho \partial_t v_i^{(\alpha)} + \partial_j \left(  \rho v_j^{(\alpha)} v_i^{(\alpha)} \right) = \partial_j \pi_{ij}^{(\alpha)},
\label{mom.eqn}
\end{equation}
where the stress tensor is
\begin{equation}
\pi_{ij}^{(\alpha)} = - P^{(\alpha)} \delta_{ij} + \sigma_{ij}^{(\alpha)},
\end{equation}
where we have separated $P^{(\alpha)}$, the hydrostatic pressure, from the 
stress $\sigma_{ij}^{(\alpha)}$. Incompressible flow is assumed in both layers
\begin{equation}
\partial_i v_i^{(\alpha)} = 0.
\label{continuity.eqn}
\end{equation}

We now consider a different viscoelastic modulus in each layer and write 
\begin{equation}
\sigma_{ij}^{(\alpha)} = \int_{-\infty}^t dt^\prime \; 
G^{(\alpha)}(t-t^\prime) \Big[ \partial_j v_i^{(\alpha)}(t^\prime) + 
\partial_i v_j^{(\alpha)}(t^\prime) \Big] .
\label{relaxation.modulus}
\end{equation} 
In order to make the problem as general as possible, we do not
specify any functional form of the moduli of the two viscoelastic fluids,
although we must point out that our study is confined to linear
viscoelasticity. We next define the complex modulus \cite{re:ferry.viscoelasticity.70}
\begin{equation}
G^{*}(\omega) = i \omega \int^{\infty}_0 d t \; G(t) e^{-i\omega t} = 
G^{\prime}(\omega) + i G^{\prime\prime}(\omega),
\label{complex.modulus}
\end{equation}
where $G^{\prime}(\omega)$ is the storage modulus, and 
$G^{\prime\prime}(\omega)$ the loss modulus.

The interface separating the two layers is $z_s = z_s(x,y,t)$. Since fluid 
particles move with the interface (no mass transport across the interface is 
allowed) we have,
\begin{equation}
\partial_t z_s + u^{(\alpha)}|_{z=z_s} \partial_x z_s = w^{(\alpha)}|_{z=z_s}.
\label{interface.eqn}
\end{equation}

The set of governing equations (\ref{mom.eqn}) - (\ref{continuity.eqn}) is to 
be supplemented by boundary conditions. We use the no-slip and no-penetration 
boundary condition on the planes. For the upper fluid,
\begin{equation}
v_i^{(2)} (z=H, t) = (U \cos(\omega t), 0, 0),
\end{equation}
and, for the lower fluid,
\begin{equation}
v_i^{(1)} (z=-h, t) = 0.
\end{equation}
At the interface ($z=z_s$) the velocity is continuous
\begin{equation}
v_i^{(2)} (z= z_s) = v_i^{(1)} (z=z_s).
\end{equation}
The stress tangential to the interface is continuous whereas the stress normal 
to the interface is discontinuous and balanced with the interfacial tension $T$
\begin{equation}
\pi_{ij}^{(2)} n_j t_i |_{z=z_s} = \pi_{ij}^{(1)} n_j t_i |_{z=z_s},
\end{equation}
\begin{equation}
\pi_{ij}^{(2)} n_j n_i |_{z=z_s} - \pi_{ij}^{(1)} n_j n_i |_{z=z_s} = 
-T(\partial_x^2 + \partial_y^2) z_s,
\label{gen.normal.stress.boundary}
\end{equation}
where $n_i$ and $t_i$ are the unit vectors normal and tangential to the 
interface. Therefore, Eqs.~(\ref{mom.eqn}) - (\ref{gen.normal.stress.boundary})
completely describe the dynamics of two superposed viscoelastic fluids under 
oscillatory shears. 

\subsection{Reference state}

In the base reference state the interface is planar ($z_s = 0$), and the flow 
is stratified and parallel to the planes, with a gradient along $\hat{z}$-axis, 
and oscillating with the driving frequency $\omega$. Hence 
${\bf v}_b^{(\alpha)} = u_b^{(\alpha)}(z,t) \hat{x}$ and we assume
\begin{equation}
u_b^{(\alpha)}(z,t) = u_{-}^{(\alpha)}(z) e^{-i\omega t} + 
u_{+}^{(\alpha)}(z) e^{i\omega t}.
\end{equation}
When ${\bf v}_b^{(\alpha)}$ is substituted into the momentum conservation 
equation, its $z$-component becomes $\partial_i P^{(\alpha)} = 0$, leading a 
constant hydrostatic pressure $P_b^{(\alpha)}$. On the other hand, the 
equation for the $x$-component involving the mode $e^{-i\omega t}$ is 
\begin{equation}
- i \omega \rho u_{-}^{(\alpha)} e^{-i\omega t} = \int_{-\infty}^t 
d t^\prime \; G^{(\alpha)}(t-t^\prime) e^{-i\omega t^\prime} 
\partial_z^2 u_{-}^{(\alpha)}.
\label{base.flow.momentum.eqn}
\end{equation}
For the $e^{i\omega t}$ mode, $u_{+}$ satisfies the complex conjugate of 
Eq.~(\ref{base.flow.momentum.eqn}); therefore, $u_{+}^{(\alpha)} = 
[u_{-}^{(\alpha)}]^{*}$ and $u_b^{(\alpha)}$ is real. With the definition of 
the complex modulus Eq.~(\ref{complex.modulus}), 
Eq.~(\ref{base.flow.momentum.eqn}) becomes
\begin{equation}
- \omega^2 \rho u_{-}^{(\alpha)} = G^{*(\alpha)}(-\omega)\partial_z^2 
u_{-}^{(\alpha)}.
\label{diff.eqn.base.flow}
\end{equation}
Note that, since $G(t)$ is real, $G^{*}(-\omega) = \Big[ G^{*}(\omega) 
\Big]^{*} = G^{\prime}(\omega) - i G^{\prime \prime}(\omega)$.

The boundary conditions for the Fourier $e^{-i\omega t}$ mode are 
$u_{-}^{(2)}(z=H) = U / 2$
and $u_{-}^{(1)}(z=-h) = 0.$ At the planar interface $n_i=(0,0,-1)$, 
$t_i=(1,0,0)$ so that continuity of velocity is $u_{-}^{(2)}(z=0) = 
u_{-}^{(1)}(z=0).$ In addition, the continuity of tangential stresses at the 
interface leads to a pressure that is continuous and constant across the 
interface $P_b^{(\alpha)} = P_b$. The balance condition 
of the normal stresses at the interface provides the following restriction 
to fluid motion in the layers
\begin{equation}
G^{*(2)}(-\omega) \partial_z u_{-}^{(2)}(z=0) = G^{*(1)}(-\omega) 
\partial_z u_{-}^{(1)}(z=0).
\label{base.flow.boundary4.1}
\end{equation}
Since the interface is not distorted in the base state, the effect of the 
surface tension is absent. 

We now solve the differential equation (\ref{diff.eqn.base.flow}) by assuming
\begin{equation}
u_{-}^{(\alpha)}(z) = A_{-}^{(\alpha)} e^{k^{(\alpha)} z} + 
B_{-}^{(\alpha)} e^{-k^{(\alpha)} z}.
\end{equation}
Upon substitution into Eq.~(\ref{diff.eqn.base.flow}), and using all four 
boundary conditions, we find
\begin{equation}
u_{-}^{(1)} (z) = \frac{U}{2K} \sinh k^{(1)} (h + z),
\label{base.flow1}
\end{equation}
\begin{equation}
u_{-}^{(2)} (z) = \frac{U}{2K} \bigg[ \sinh k^{(1)} h \cosh k^{(2)} z + 
\frac{G^{*(1)}(-\omega) k^{(1)}}{G^{*(2)}(-\omega) k^{(2)}} \cosh k^{(1)} h 
\sinh k^{(2)} z \bigg],
\label{base.flow2}
\end{equation}
where
\begin{equation}
k^{(\alpha)} = i \sqrt{\frac{\omega^2 \rho}{G^{*(\alpha)}(-\omega)}},
\label{base.flow.wavevector}
\end{equation}
and
\begin{equation}
K = \sinh k^{(1)} h \cosh k^{(2)} H + \frac{G^{*(1)}(-\omega) 
k^{(1)}}{G^{*(2)}(-\omega) k^{(2)}} \cosh k^{(1)} h \sinh k^{(2)} H.
\end{equation}
Note that the reference state explicitly depends on the complex moduli at the 
driving frequency $\omega$. In the next section we study the linear stability 
of this base flow against a small perturbation of the interface.

\section{Linear Stability Analysis}
\label{sec:linear}

We introduce small perturbations away from the base state ${\bf v} = {\bf v}_b 
+ \delta {\bf v}^{(\alpha)}$, with $\delta {\bf v}^{(\alpha)} = 
(\delta u^{(\alpha)}, 0, \delta w^{(\alpha)})$, and 
$ P^{(\alpha)} = P_b + \delta P^{(\alpha)}.$ Since our system is effectively 
two dimensional, we introduce the stream function $\phi^{(\alpha)}$ such 
that $\delta {u}^{(\alpha)} = \partial_z \phi^{(\alpha)}$ and 
$\delta {w}^{(\alpha)} = - \partial_x \phi^{(\alpha)}$.
We decompose the perturbations into normal modes in the $x$ direction
\begin{equation}
\phi^{(\alpha)} =  \hat{\phi}^{(\alpha)}(z,t) e^{i q_x x}, \quad 
\delta P^{(\alpha)} = \delta \hat{P}^{(\alpha)} (z, t)  e^{i q_x x}
\end{equation}
and
\begin{equation}
z_s = \hat{z}_s (t)  e^{i q_x x}
\end{equation}
In general, $\hat{\phi}^{(\alpha)}(z,t)$, $\hat{z}_s (t)$ and 
$\delta \hat{P}^{(\alpha)} (z, t)$ are complex and depend on $q_x$. The 
resulting differential equations, after linearizing in the amplitude of the 
perturbations, have time periodic coefficients in $2\pi/\omega$, and hence 
we use Floquet theory. We therefore write 
\begin{equation}
\hat{\phi}^{(\alpha)} = \bar{\phi}^{(\alpha)}(z,t) e^{\sigma t}, \quad \delta 
\hat{P}^{(\alpha)} = \delta \bar{P}^{(\alpha)}(t) e^{\sigma t},
\label{floquet.perturbation}
\end{equation}
and
\begin{equation}
\hat{z}_s^{(\alpha)} = \bar{z}^{(\alpha)}_s(t) e^{\sigma t},
\label{floquet.perturbation2}
\end{equation}
where $\bar{\phi}^{(\alpha)}(z,t)$, $\bar{z}_s^{(\alpha)}(t)$ and 
$\delta \bar{P}^{(\alpha)}(t)$ are periodic functions of time with period 
$2\pi/\omega$, and $\sigma$ is the Floquet exponent which determines stability.
When $\Re[\sigma]$ becomes positive, perturbations grow exponentially and the 
system becomes unstable. The remainder of this paper concerns the calculation 
of the Floquet exponent.

Upon applying Floquet theory, the resulting equations of motion in terms of 
the stream function $\bar{\phi}^{(\alpha)}$ and the interface $\bar{z}_s$ are
\begin{equation}
\begin{split}
\rho (\partial_t + \sigma) (\partial_z^2 - &q_x^2) \bar{\phi}^{(\alpha)} + 
i \rho q_x  u_b^{(\alpha)} (\partial_z^2 - q_x^2) \bar{\phi}^{(\alpha)} - 
i \rho q_x \bar{\phi}^{(\alpha)} (\partial_z^2 u_b^{(\alpha)})
\\
&= \int_{-\infty}^t d t^\prime \; G^{(\alpha)}(t-t^\prime) (\partial_z^2 - 
q_x^2)^2 \bar{\phi}^{(\alpha)}  e^{\sigma (t^\prime - t)},
\label{2D.lin.OS.eqn.1}
\end{split}
\end{equation}
and 
\begin{equation}
(\partial_t + \sigma) \bar{z}_s + i q_x u_b^{(\alpha)}|_{z=0} \bar{z}_s = 
- i q_x \bar{\phi}^{(\alpha)}|_{z=0}.
\label{interface.motion}
\end{equation}
In addition, the no-slip and no-penetration boundary conditions at the planes 
become
\begin{equation}
\bar{\phi}^{(2)} (z=H) =
\partial_z \bar{\phi}^{(2)} (z=H) =
\bar{\phi}^{(1)} (z=-h) =
\partial_z \bar{\phi}^{(1)} (z=-h) = 0.
\label{2D.boundary.cond1}
\end{equation}
and the interface conditions are
\begin{equation}
\bar{\phi}^{(2)}(z=0) = \bar{\phi}^{(1)}(z=0),
\end{equation}
and 
\begin{equation}
\partial_z u_b^{(2)}(z=0) \bar{z}_s + \partial_z \bar{\phi}^{(2)}(z=0) = 
\partial_z u_b^{(1)}(z=0) \bar{z}_s + \partial_z \bar{\phi}^{(1)}(z=0).
\label{2D.interface.cond2}
\end{equation}
For small $z_s$ we have $n_i = (\partial_x z_s, 0, -1)$ and  
$t_i = (1,0,\partial_x z_s)$. Then, the continuity of the tangential shear 
stresses implies that
\begin{equation}
\delta \bar{\sigma}_{xz}^{(2)}|_{z=0} = \delta \bar{\sigma}_{xz}^{(1)}|_{z=0},
\end{equation}
where 
\begin{equation}
\delta \bar{\sigma}_{xz}^{(\alpha)} =
\int_{-\infty}^t d t^\prime \; G^{(\alpha)}(t - t^\prime) 
( \partial_z^2 + q_x^2) \bar{\phi}^{(\alpha)}(t^\prime)  
e^{\sigma (t^\prime - t)}.
\end{equation}
Note that the pressure term in the stress tensor vanishes because 
$\delta_{ij} n_i t_j = 0$, and the contribution of the base flow to this 
boundary condition is not present because the base flow is continuous at 
$z=0$, and there is no density difference between two fluids. If the 
densities of two fluids are different, one cannot neglect the contribution 
of base flow. On the other hand, the normal component of the stress tensor 
must be balanced with the surface tension as
\begin{equation}
\Big[ - i q_x \delta \bar{P}^{(2)} + i q_x \delta \bar{\sigma}_{zz}^{(2)} 
\Big]_{z=0} - \Big[ - i q_x \delta \bar{P}^{(1)} + i q_x \delta 
\bar{\sigma}_{zz}^{(1)} \Big]_{z=0} = i q_x^3 T \bar{z}_s,
\label{2D.normal.stress}
\end{equation}
where
\begin{eqnarray}
- i q_x \delta \bar{P}^{(\alpha)}  +i q_x \delta \bar{\sigma}_{zz}^{(\alpha)}
&=& \rho (\partial_t + \sigma) \partial_z \bar{\phi}^{(\alpha)} + 
i \rho q_x u_b^{(\alpha)} \partial_z \bar{\phi}^{(\alpha)} - 
i \rho q_x \bar{\phi}^{(\alpha)} (\partial_z u_b^{(\alpha)})
\nonumber
\\
&&
- \int_{-\infty}^t d t^\prime \; G^{(\alpha)}(t-t^\prime) (\partial_z^2 - 
3 q_x^2) \partial_z \bar{\phi}^{(\alpha)}   e^{\sigma (t^\prime - t)}.
\end{eqnarray}

In order to make analytic progress, we focus only on long-wavelength 
perturbations. For small $q_x$, we expand the perturbations equations 
(\ref{floquet.perturbation}) and (\ref{floquet.perturbation2}) in $q_x$ 
such that
\begin{equation}
\bar{\phi}^{(\alpha)}(z,t) = \bar{\phi}^{(\alpha)}_0(z,t) + 
\bar{\phi}^{(\alpha)}_1(z,t) q_x + \bar{\phi}^{(\alpha)}_2(z,t) q_x^2 + \ldots,
\end{equation}
\begin{equation}
\bar{z}_s(t) = \bar{z}_{s, 0}(t) + \bar{z}_{s, 1}(t) q_x + 
\bar{z}_{s, 2}(t) q_x^2 + \ldots,
\end{equation}
\begin{equation}
\sigma = \sigma_0 + \sigma_1 q_x + \sigma_2 q_x^2 + \ldots,
\end{equation}
where at each order in $q_x$ the coefficients of $\bar{\phi}^{(\alpha)}(z,t)$ 
and $\bar{z}_s(t)$ are time-periodic functions of $2\pi / \omega$, whereas 
the coefficients of $\sigma$ are independent of time. Next, after 
substituting these perturbations into Eqs.~(\ref{2D.lin.OS.eqn.1}) 
and (\ref{interface.motion}) we solve the set of dynamic equations order by order in $q_x$. 

Before delving into the calculation, we can determine first from the equation 
of motion of the interface, Eq.~(\ref{interface.motion}), at which order in 
$q_x$ the first non-vanishing Floquet exponent appears, and its functional 
form in the base flow $u_b$ and the perturbed stream function. 
At $\mathcal{O}(1)$ in $q_x$ the equation of motion of the interface becomes
\begin{equation}
\sigma_0 \bar{z}_{s,0} + \partial_t \bar{z}_{s,0} = 0
\label{2D.interface.eqn.1}
\end{equation}
According to Floquet theory $\bar{z}_{s,0}$ must be a periodic function of time. 
Then it follows from this equations that $\sigma_0 = 0$ and $\bar{z}_{s,0} = 
\zeta_0$, constant. 

At $\mathcal{O}(q_x)$, with $\sigma_0 = 0$ and $\bar{z}_{s,0} = \zeta_0$, we 
have
\begin{equation}
\sigma_1 \zeta_0 + \partial_t \bar{z}_{s,1} +i u_b|_{z=0} \zeta_0 = 
-i \bar{\phi}_0|_{z=0}.
\label{sigma.q1}
\end{equation}
The time-dependence of $\bar{\phi}_{0}^{(\alpha)}(t)$ can be inferred from 
the continuity condition of the $\hat{x}$ component velocity at the interface 
Eq.~(\ref{2D.interface.cond2}). Since $\bar{z}_{s,0} = \zeta_0$ and 
$u_b \sim e^{\pm i \omega t}$, $\bar{\phi}_0$ is also periodic in time, and we 
take
\begin{equation}
\bar{\phi}_{0}^{(\alpha)}(z,t) = \bar{\phi}_{0,-}^{(\alpha)}(z) 
e^{-i\omega t} + \bar{\phi}_{0,+}^{(\alpha)}(z) e^{i\omega t}.
\label{phi0fnform}
\end{equation}
Then, since $\bar{z}_{s,1}$ must be a periodic function in time because of Floquet theory, 
it should follow again from Eq.~(\ref{sigma.q1}) that $\sigma_1 = 0$ and 
\begin{equation}
\partial_t \bar{z}_{s,1} = - i \Big( u_b|_{z=0} \zeta_0 + \bar{\phi}_0|_{z=0} 
\Big),
\end{equation}
from which we find
\begin{equation}
\bar{z}_{s,1} =
\frac{1}{\omega} \Big[ \zeta_0 u_{-}^{(\alpha)} + 
\bar{\phi}_{0,-}^{(\alpha)} \Big]_{z=0} e^{-i\omega t} - \text{c.c.},
\label{zs1}
\end{equation}
where c.c. stands for complex conjugation. Note that $[\bar{z}_{s,1}]^{*} = 
- \bar{z}_{s,1}$. In calculating $\bar{z}_{s,1}$ it does not matter whether 
we use the base flow and stream function of fluid 1 or the fluid 2 because 
of the continuity condition at the interface. Since $\sigma_1=0$, the first 
non-vanishing contribution to the Floquet exponent appears at 
$\mathcal{O}(q_x^2)$. 

Consider then the problem at $\mathcal{O}(q_x^2)$, with $\sigma_{0} = \sigma_{1} 
= 0$ and $\bar{z}_{s,0} = \zeta_0$. We have 
\begin{equation}
\sigma_2 \zeta_0 + \partial_t \bar{z}_{s,2} = - i u_b^{(\alpha)}|_{z=0} 
\bar{z}_{s,1} -i \bar{\phi}_1^{(\alpha)}|_{z=0}.
\label{sigma.q2}
\end{equation}
Similarly to the case at $\mathcal{O}(q_x)$ the time-dependence of 
$\bar{\phi}_1^{(\alpha)}$ is determined from the boundary condition 
Eq.~(\ref{2D.interface.cond2}). In this case, since $\bar{z}_{s,1}$ is a 
periodic function of time rather than a constant, and $\partial_z 
u_b^{(\alpha)}(z=0) \bar{z}_{s,1}$ has both a non periodic part and a time 
periodic part, the perturbed stream function at $\mathcal{O}(q_x)$ is also a 
sum of non-periodic and periodic parts as $\bar{\phi}_1^{(\alpha)} = 
\bar{\phi}_{1,\text{periodic}}^{(\alpha)} + 
\bar{\phi}_{1,\text{NP}}^{(\alpha)}$. Thus, from Eq.~(\ref{sigma.q2}) we find 
the periodic part
\begin{equation}
\partial_t \bar{z}_{s,2} = -i \Big[ u_b^{(1)}|_{z=0} \bar{z}_{s,1} + 
\bar{\phi}_1^{(1)}|_{z=0} \Big]_\text{periodic},
\end{equation}
and the first non-vanishing contribution to the Floquet exponent is
$\sigma = \sigma_2 q_x^2$ where
\begin{equation}
\sigma_2=
- \frac{i}{\zeta_0} \Big[ u_b^{(1)}|_{z=0} \bar{z}_{s,1} + 
\bar{\phi}_1^{(1)}|_{z=0} \Big]_\text{NP},
\label{sigma2}
\end{equation}
with
\begin{equation}
\Big[ u_b^{(1)}|_{z=0} \bar{z}_{s,1} \Big]_\text{NP} = \frac{1}{\omega} 
\Big[ u_{+}^{(1)} \bar{\phi}_{0,-}^{(1)} - \text{c.c} \Big]_{z=0}.
\end{equation}
Therefore, in order to determine the stability of the configuration under 
study, we need the first order distortion of interface $\bar{z}_{s,1}$ and the 
non-periodic part of $\bar{\phi}_1^{(\alpha)}(z=0)$.

We show the details of the calculation of the required contributions in 
Appendix~\ref{sec:appendixA}. A complete analytic solution can be found in 
the limits of small Reynolds number and small elasticity. We mention that a 
complete analytic solution for arbitrary Reynolds number can also be found; 
however, it is too complicated to present here, and the Floquet eigenvalue 
needs to be evaluated numerically. As the limit of relevance in the copolymer 
case is that of vanishing Reynolds number, we do not pursue this general case 
here. Nevertheless, Appendix~\ref{sec:appendixB} describes the steps necessary 
to find the Floquet eigenvalue for arbitrary Reynolds numbers and elasticity. 

For convenience, we define a complex Reynolds number
\begin{equation}
cRe^{(\alpha)} = \frac{\rho \omega^2 h^2}{G^{*(\alpha)}(\omega)} = 
-i \frac{Re^{(\alpha)}}{1 - i g^{(\alpha)}},
\end{equation}
where $g^{(\alpha)} = G^{\prime (\alpha)}(\omega) / G^{\prime \prime 
(\alpha)}(\omega)$, and $Re^{(\alpha)} = \rho \omega^2 h^2 / G^{\prime 
\prime (\alpha)}(\omega)$ are the conventional Reynolds numbers of fluids 1 
and 2 with $Re = Re^{(1)}$. In addition, we define $n^{(\alpha)} =  
G^{\prime \prime (1)}(\omega) / G^{\prime \prime (\alpha)}(\omega)$ with 
$n = n^{(2)}$. The limit of small Reynolds number implies that both 
$Re^{(1)}$ and $Re^{(2)} H^2/h^2$ are small. Now we solve the equations of 
motion (\ref{2D.lin.OS.eqn.1}) for the stream function order by order in 
$q_x$ by taking the expansion in $Re$ and $g^{(\alpha)}$ as
\begin{equation}
\bar{\phi}_{0,-}^{(\alpha)}(z) = \bar{\phi}_{0,-,0}^{(\alpha)}(z)+ 
Re \bar{\phi}_{0,-,1}^{(\alpha)}(z) + \mathcal{O}(Re^2, (g^{(\alpha)})^2),
\label{eq:form.phi0}
\end{equation}
and
\begin{equation}
\bar{\phi}_{1,\text{NP}}^{(\alpha)}(z) = 
\bar{\phi}_{1,\text{NP},0}^{(\alpha)}(z)+ 
Re \bar{\phi}_{1,\text{NP},1}^{(\alpha)}(z) + \mathcal{O}(Re^2, (g^{(\alpha)})^2),
\label{eq:form.phi1NP}
\end{equation}
where $\bar{\phi}_{0,-,0}^{(\alpha)}(z)$ and 
$\bar{\phi}_{1,\text{NP},0}^{(\alpha)}(z)$ are linearly proportional to 
$g^{(\alpha)}$ and $\bar{\phi}_{0,-,1}^{(\alpha)}(z)$ and 
$\bar{\phi}_{1,\text{NP},1}^{(\alpha)}(z)$ do not depend on $g^{(\alpha)}$ 
(See Appendix~\ref{sec:appendixA} for details). 
For long wavelength perturbations it is found that the surface tension $T$ 
does not enter in the calculation because it appears at third order in $q_x$.

\begin{figure}[t]
\centering
  \subfigure[$n<1$]
  {
    \centering
\includegraphics[width=0.5\textwidth]{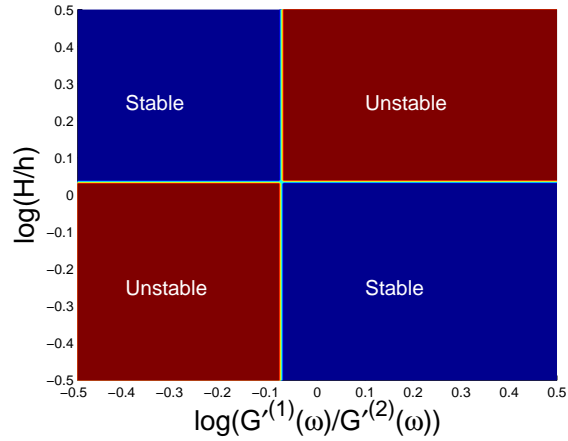}
    \label{fig2a.ins1}
  }
~
  \subfigure[$n>1$]
  {
    \centering
\includegraphics[width=0.5\textwidth]{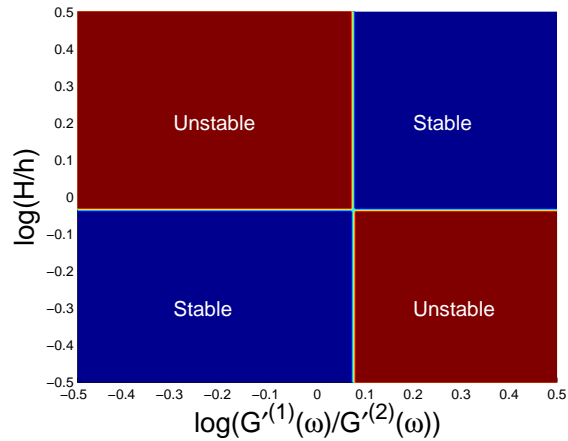}
    \label{fig2b.ins2}
  }
\caption{Stability Diagram of the two layer configuration. 
The vertical neutral stability line is set by 
$G^{\prime (1)}(\omega) = n G^{\prime (2)}(\omega)$ 
while the horizontal line $h = \sqrt{n} H$. 
For example, we have taken: a) $n = 5/6$ and b) $n= 6/5$.}
\label{fig2}
\end{figure}

From Eq.~(\ref{sigma2}) with Eqs.~(\ref{A0F1}), (\ref{E0F1}), (\ref{A1F1}) 
and (\ref{E1F1}), we find for small $Re$ and $g^{(\alpha)}$
\begin{equation}
\sigma_2 = \sigma_\text{el} + \sigma_\text{vis} Re + 
\mathcal{O}(Re^2,(g^{(\alpha)})^2),
\label{eq:gen.sigma2}
\end{equation}
%
%
where the explicit expressions of $\sigma_\text{el}$ and $\sigma_\text{vis}$ are given 
in Eqs.~(\ref{eq:sigma2.elastic}) and (\ref{eq:sigma2.viscous}). 
The term $\sigma_{\text{el}}$ that remains as $Re \rightarrow 0$ is a 
contribution of elastic origin. We find that $\sigma_{\text{el}} \propto 
(n-1) (nH^2 - h^2) (g^{(2)} - g^{(1)})$, where the proportionality factor is 
positive. This term vanishes if either the elasticity ($g^{(2)} = g^{(1)}$) 
or viscosity ($n = 1$) of both fluids is the same. 
In short, we find a long wavelength instability due to elasticity 
stratification when $(n-1) (nH^2 - h^2) (g^{(2)} - g^{(1)}) > 0$. 
Figure \ref{fig2} shows a typical stability diagram.

In addition, there is a contribution purely of viscous origin that appears 
linearly in $Re$, $\sigma_\text{vis}$. It agrees with the earlier calculation 
of Ref.~\onlinecite{huang07} in which the constitutive relation of the phases 
was assumed to be uniaxial but Newtonian. In this latter case, a long 
wavelength interfacial instability occurs whenever the thinner layer is 
more viscous.

An important difference between our calculation and the case of Newtonian 
layers is the fact that here the up-down symmetry of the configuration is 
lost: $\sigma_\text{el}$ changes sign upon reversing the position of the 
fluid layers and their thicknesses. The sign of $\sigma_\text{el}(H,h,G^{*(2)}(\omega), 
G^{*(1)}(\omega))$ is opposite to the sign of $\sigma_\text{el}(h,H,G^{*(1)}(\omega),
%
%
G^{*(2)}(\omega))$. Therefore, stability depends on which layer is adjacent 
to the boundary being sheared. 

%
%
The instability mechanism in the Newtonian case arises from the out of phase
evolution of the vorticity in both fluids \cite{hinch84, king99}. Small periodic
distortions of the interface induce a vorticity perturbation of the same sign
across the interface, but its sign alternates at each trough and peak.  With
negligible inertia, the vorticities are advected by the shear to create out of
phase components of vorticity because of the viscosity contrast. As a result,
vorticities of opposite sign at adjacent troughs and peaks get close together,
and induce a vertical motion of the interface.  The interfacial instability due
to elasticity contrast found here has a similar origin in the out of phase
vorticity as in the viscous case. The requirement that a viscosity contrast must
exist ($n \neq 1$) for instability to occur ensures that there should exist an
imbalance of vorticity across the interface.  The essential difference between
the viscous and elastic cases is that the out of phase motion of vorticity is
driven in our case by the elasticity response.  As can be seen from the complex
wavevector of the base state, Eq.~(\ref{base.flow.wavevector}), elasticity
induces a phase shift relative to the driving shear flow.  Similarly, the
elastic contrast of two fluids generates out of phase components of perturbed
vorticity, and drives the interfacial instability.

\section{Discussion}
\label{sec:discussion}

%
%


%
%

We discuss in this section the possible implications of our findings on
orientation selection of block copolymers under shear. Following a quench of a
large aspect ratio configuration (the aspect ratio being the ratio between the
lateral dimension of the system and the lamellar spacing), a spontaneous
distribution of locally ordered lamellar domains with random orientation results
that coarsens with time \cite{re:elder92}. Under the application of a shear, the
wavelength of those domains with orientation that has a transverse component
($\mathbf{\hat{q}} \parallel \mathbf{v}$) changes, fact that greatly reduces
their range of stability \cite{re:chen02}. Therefore, after some initial
transient, it is expected that a majority of domains would be oriented either
parallel or perpendicular to the shear. Within the order parameter description
used here, the local free energy of domains along the two orientations is the
same, even under shear (since $\mathbf{\hat{q}} \cdot \mathbf{v} = 0$). Hence,
they are in coexistence.

As noted above, however, viscosity or elasticity of A or B rich regions are
typically different. In PS-PI the glass transition temperature of the two blocks
is quite different, and hence their mechanical response within the copolymer is
expected to be different. However, given the small size of the blocks (on the
order of tens of nm) and the large average viscosity of polymer melts, flows at
the lamellar scale are expected to be negligible under most conditions of
interest \cite{re:tamate08}. Furthermore, their contribution in slightly
distorted lamellar configurations is of second order in the distortion, and do
not contribute appreciably to their relaxation or stability \cite{re:chen02}.
Therefore viscoleastic contrast between blocks seems insufficient to drive short
scale flows that could account for differences between parallel and
perpendicular orientations under shear.

Whereas short scale flows are strongly damped, this is not so for flows that
couple to long range distortions of a lamellar configuration \cite{yoo12}, or
flows at the scale of the characteristic domain size in a polycrystalline
configuration. The analysis undertaken here has aimed at establishing the
relative dynamic stability of parallel and perpendicular domains when long
wavelength hydrodynamic flows are allowed. We have found that there exists an
interfacial instability of hydrodynamic origin in that limit due to viscoelastic
contrast between the two phases even for zero Reynolds number.
%
%



\begin{figure}[t] \centering
\includegraphics[width=12cm]{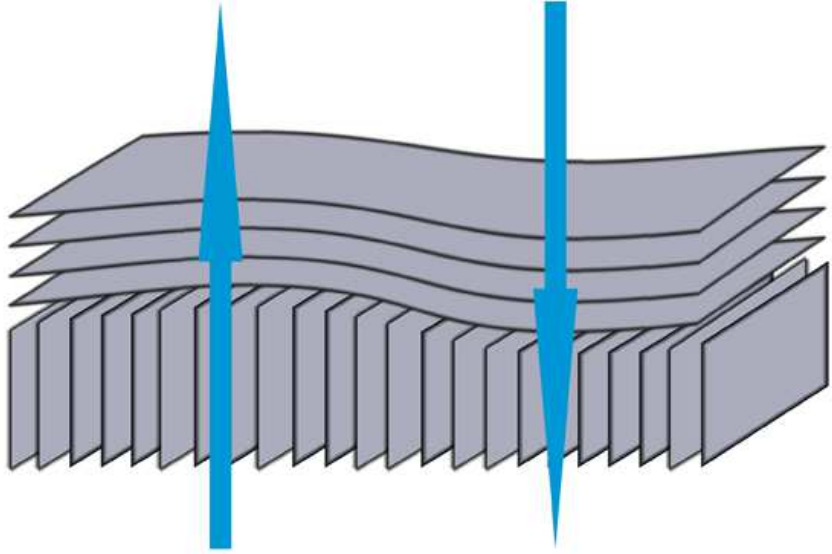}
\caption{Orientation selection. The secondary flow induced by the unstable
configuration is indicated by light blue arrows. This flow distorts the 
parallel lamellae without affecting the perpendicular lamellae.} 
\label{fig3}
\end{figure}

Domain coarsening and orientation selection are intrinsically nonlinear problems
whereas we have only addressed here the linear stability of a particular
configuration. The two can be qualitatively related as follows: Assume that
after some transient there will be a preponderance of parallel and
perpendicular domains in a large aspect ratio sample under oscillatory shear.
The instability described will set in whenever two such
domains meet, and the conditions for instability are satisfied (viscoelastic 
contrast) and low enough wavelength (or equivalently, a large enough 
characteristic domain size).
Once any boundary between two such domains becomes unstable, a secondary flow 
is established which is normal to the boundary (shown schematically in 
Fig.~\ref{fig3}). Advection of order parameter by this secondary flow distorts 
the parallel domain ($\mathbf{v} \parallel \mathbf{q}$) but not the 
perpendicular ($\mathbf{v} \perp \mathbf{q}$). The region with locally 
distorted parallel lamellae will have a higher 
free energy than the adjacent region of perpendicular lamellae, and 
boundary motion will follow to reduce the free energy imbalance. As a 
consequence, we would expect systematic motion of the boundary towards the 
parallel region every time there is a boundary instability. Although we cannot 
predict the nonlinear evolution of the boundary, and hence the evolution of 
an ensemble of domains in a large aspect ratio sample, this instability can
provide for a dynamical selection mechanism that favors the perpendicular 
orientation whenever there is viscoelastic contrast between the two 
orientations. This result is generally
consistent with the experimental survey of Ref. \onlinecite{wu05}.

Three main issues remain unresolved. First, the argument presented does not
account for the preponderance of the parallel orientation when there is no
contrast. Second, as shown in Fig.~\ref{fig2},
there are regions of parameter space in which the parallel-perpendicular 
boundary is stable. Therefore, whether a distribution of domains in a large
aspect ratio sample would ultimately coarsen or evolve into coexistence cannot 
be conclusively answered by our analysis. Third, although there is ample
evidence of viscoelastic contrast between differently oriented lamellae, 
there is no experimental study that we are aware of that has directly 
probed the uniaxial hydrodynamic properties of a bulk lamellar phase.

\begin{acknowledgements}
We are indebted to Zhi-Feng Huang for many useful conversations, and thank
the Minnesota Supercomputing Institute for support.  
\end{acknowledgements}

\bibliographystyle{prsty}
\bibliography{interface}

\appendix

\section{Expansion results for low Reynolds number and small elasticity}
\label{sec:appendixA}

\subsection{$\mathcal{O}(1)$ in $q_x$}

At zeroth order in $q_x$ the momentum conservation equation
Eq.~(\ref{2D.lin.OS.eqn.1}) reduces to, with $\sigma_0 = 0$, 
\begin{equation}
\rho \partial_t \partial_z^2 \bar{\phi}_0^{(\alpha)} = \int_{-\infty}^t d
t^\prime \; G^{(\alpha)}(t-t^\prime) \partial_z^4 \bar{\phi}_0^{(\alpha)}.
\label{phi0.O1} 
\end{equation} 
We have shown earlier that
$\bar{\phi}_0^{(\alpha)}$ is time-periodic in $2\pi/\omega$, and given by
Eq~(\ref{phi0fnform}). For $e^{- i\omega t}$ mode, we have 
\begin{equation}
-\frac{[cRe^{(\alpha)}]^{*}}{h^2} \partial_z^2 \bar{\phi}_{0,-}^{(\alpha)} =
\partial_z^4 \bar{\phi}_{0,-}^{(\alpha)}.  \end{equation} For small $Re$, with
Eq.~(\ref{eq:form.phi0}), the above equation (\ref{phi0.O1}) becomes at zeroth
order in $Re$ 
\begin{equation} 
\partial_z^4 \bar{\phi}_{0,-,0}^{(\alpha)}(z) =
0, 
\label{diff1} 
\end{equation} 
and at first order in $Re$ 
\begin{equation}
\partial_z^4 \bar{\phi}_{0,-,1}^{(\alpha)} = -\frac{i}{h^2}
\frac{n^{(\alpha)}}{1+ig^{(\alpha)}} \partial_z^2 \bar{\phi}_{0,-,0}^{(\alpha)}.
\label{diff2} 
\end{equation} 
First, the general solution of Eq.~(\ref{diff1}) is
\begin{equation} 
\bar{\phi}_{0,-,0}^{(\alpha)}(z) = A_0^{(\alpha)} +
B_0^{(\alpha)} z + C_0^{(\alpha)} z^2 + D_0^{(\alpha)} z^3. 
\label{gen.sol.1}
\end{equation} 
The coefficients are determined by applying the boundary
conditions Eqs.~(\ref{2D.boundary.cond1}) - (\ref{2D.normal.stress}). At 
zeroth order in $Re$ the boundary conditions are
\begin{equation}
\bar{\phi}^{(2)}_{0,-,0} (z=H) = 0, 
\end{equation} 
\begin{equation} 
\partial_z
\bar{\phi}_{0,-,0}^{(2)} (z=H) = 0, 
\end{equation} 
\begin{equation}
\bar{\phi}_{0,-,0}^{(1)} (z=-h) = 0, 
\end{equation} 
\begin{equation} 
\partial_z
\bar{\phi}_{0,-,0}^{(1)} (z=-h) = 0, 
\end{equation} 
\begin{equation}
\bar{\phi}^{(2)}_{0,-,0}(z=0) = \bar{\phi}^{(1)}_{0,-,0}(z=0), 
\end{equation}
\begin{equation} 
\partial_z u_{-,0}^{(2)}(z=0) \bar{z}_{s,0} + \partial_z
\bar{\phi}^{(2)}_{0,-,0}(z=0) = \partial_z u_{-,0}^{(1)}(z=0) \bar{z}_{s,0} +
\partial_z \bar{\phi}^{(1)}_{0,-,0}(z=0), 
\end{equation} 
\begin{equation}
G^{*(2)}(-\omega) \partial_z^2 \bar{\phi}_{0,-,0}^{(2)} (z=0) =
G^{*(1)}(-\omega) \partial_z^2 \bar{\phi}_{0,-,0}^{(1)} (z=0), 
\end{equation}
\begin{equation} 
G^{*(2)}(-\omega) \partial_z^3 \bar{\phi}_{0,-,0}^{(2)} (z=0) =
G^{*(1)}(-\omega) \partial_z^3 \bar{\phi}_{0,-,0}^{(1)} (z=0), 
\end{equation}
where the base flow is also expanded in $Re$ 
\begin{equation} u_{-}^{(\alpha)}
(z) = u_{-,0}^{(\alpha)} (z) + u_{-,1}^{(\alpha)} (z) Re + \mathcal{O}(Re^2,
(g^{(\alpha)})^2).  
\end{equation} 
In terms of the coefficients of
$\bar{\phi}_{0,-,0}^{(\alpha)}$ this set of boundary condition can be 
expressed as a 8$\times$8 matrix. We solve this linear system for the
coefficients of $\bar{\phi}_{0,-,0}^{(\alpha)}(z)$ for small elasticity
$g^{(\alpha)} \ll 1$, obtaining 
\begin{eqnarray} 
A_0^{(1)} &=& A_0^{(2)}
\nonumber \\ 
&=& \frac{(H+h) h^2 (n-1) n H^2 \zeta_{0} U}{(h+ nH)(n^2 H^4 + 4 n
H^3 h + 6 n H^2 h^2 + 4 n h^3 H + h^4)} \nonumber \\ 
&& + i \frac{n U (H+h)
\zeta_{0} H^2 h^2 (g^{(2)} - g^{(1)})}{(h+ nH)^2(n^2 H^4 + 4 n H^3 h + 6 n H^2
h^2 + 4 n h^3 H + h^4)^2} \times \nonumber \\ 
&& \times (n^4 H^5 - 2 n^3 H^5 -5
H^4 n^2 h - 10 n^2 H^3 h^2 - 10 n^2 H^2 h^3 \nonumber \\ 
&&
- 5 n^2 H h^4 - 2 h^5 n + h^5) 
\label{A0F1} 
\end{eqnarray} 
\begin{eqnarray}
B_0^{(1)} &=& \frac{(4nH^3 + h^3 + 3 nH^2 h) (n-1) h \zeta_{0} U}{2(h+ nH)(n^2
H^4 + 4 n H^3 h + 6 n H^2 h^2 + 4 n h^3 H + h^4)} \nonumber \\ 
&& + i \frac{n U
\zeta_{0} h (g^{(2)} - g^{(1)})}{2(h+ nH)^2(n^2 H^4 + 4 n H^3 h + 6 n H^2 h^2 +
4 n h^3 H + h^4)^2} \times \nonumber \\ 
&& \times (- 5 H h^7 - 3 H^2 h^6 + 4 n^4
H^8 - h^8 - 73 n^2 H^5 h^3 + 2 n^3 H^5 h^3 \nonumber \\ 
&& -20n H^3 h^5 - 6 n^3
H^7 h -10H^4nh^4 - 55H^6n^2h^2 - 45H^4n^2h^4 \nonumber \\ 
&& +4h^6n^2H^2 -
14h^6nH^2 - 9n^2H^3h^5 + 3n^4H^7h - 20hH^7n^2 - 8n^3H^8) 
\label{B0F1}
\end{eqnarray} 
\begin{eqnarray} 
B_0^{(2)} &=& \frac{(H^3n +4h^3 + 3h^2H) (n-1) n
H \zeta_{0} U}{2(h+ nH)(n^2 H^4 + 4 n H^3 h + 6 n H^2 h^2 + 4 n h^3 H + h^4)}
\nonumber \\ 
&& + i \frac{n U H \zeta_{0} (g^{(2)} - g^{(1)})}{2(h+ nH)^2(n^2
H^4 + 4 n H^3 h + 6 n H^2 h^2 + 4 n h^3 H + h^4)^2} \times \nonumber \\ 
&&
\times (n^4 H^8 + 5 n^4 H^7 h + 3 n^4 H^6 h^2 + 14 n^3 H^6 h^2 + 20 n^3 H^5 h^3
\nonumber \\ 
&& + 10 n^3 H^4 h^4 - 4 n^2 H^6 h^2 + 9 n^2 h^5 h^3 + 45 n^2 H^4
h^4 + 73 n^2 H63 h^5 \nonumber \\ 
&& + 55 n^2 H^2 h^6 + 20 n^2 H h^7 - 2n H^3
h^5 + 6 n H h^7 + 8 n h^8 - 3 H h^7 - 4h^8) 
\label{B0F2} 
\end{eqnarray}
\begin{eqnarray} 
C_0^{(1)} &=& \frac{1}{n} \Big[1+i(g^{(2)}-g^{(1)}) \Big]
C_0^{(2)} \nonumber \\ 
&=& \frac{(H^3 n + h^3) (n-1) \zeta_{0} U}{(h+ nH)(n^2
H^4 + 4 n H^3 h + 6 n H^2 h^2 + 4 n h^3 H + h^4)} \nonumber \\ 
&& + i \frac{n U
\zeta_{0} (g^{(2)} - g^{(1)})}{(h+ nH)^2(n^2 H^4 + 4 n H^3 h + 6 n H^2 h^2 + 
4 n h^3 H + h^4)^2} \times \nonumber \\ 
&& \times (H^8 n^4 -2 n^3 H^8 +2 H^5 n^3 h^3
-5 h H^7 n^2 -10 H^6 n^2h^2 \nonumber \\ 
&& -13 H^5 n^2 h^3 +6 n^2 H^3 h^5 +4
h^6 n^2 H^2 -10 H^4 n h^4 -14 H^3 n h^5 \nonumber \\ 
&& -8 h^6 n*H^2 -3 h^5 H^3
-6 h^6 H^2 -5 h^7 H -h^8) 
\label{C0F1} 
\end{eqnarray} 
\begin{eqnarray} 
D_0^{(1)}
&=& \frac{1}{n} \Big[1+i(g^{(2)}-g^{(1)}) \Big] D_0^{(2)} \nonumber \\ 
&=&
\frac{(h^2 - n H^2) (n-1) \zeta_{0} U}{2(h+ nH)(n^2 H^4 + 4 n H^3 h + 6 n H^2
h^2 + 4 n h^3 H + h^4)} \nonumber \\ 
&& + i \frac{n U \zeta_{0} (g^{(2)} -
g^{(1)})}{2(h+ nH)^2(n^2 H^4 + 4 n H^3 h + 6 n H^2 h^2 + 4 n h^3 H + h^4)^2}
\times \nonumber \\ 
&& \times (2 n^3 H^7 -12 h^4 H^3 n +4 h^5 n^2 H^2 +5 n^2 H^6
h -7 h^5 H^2 -5 h^6 H - h^7 \nonumber \\ 
&& -6 h^5 n H^2 +15 h^3 H^4 n^2 +11 h^4
H^3 n^2 +2 h^2 n^3*H^5 +7 n^2 H^5 h^2 \nonumber \\ 
&& -10 n H^4 h^3 -n^4 H^7 -4
H^3 h^4) \label{D0F1} 
\end{eqnarray}

At $\mathcal{O}(Re)$ the general solution to Eq.~(\ref{diff2}) is given by
\begin{equation} 
\bar{\phi}_{0,-,1}^{(\alpha)}(z) = E_0^{(\alpha)} +
F_0^{(\alpha)} z + J_0^{(\alpha)} z^2 + L_0^{(\alpha)} z^3 - \frac{i}{4 h^2}
\frac{n^{(\alpha)}}{1+ig^{(\alpha)}}  \left( \frac{C_0^{(\alpha)}}{3} z^4 +
\frac{D_0^{(\alpha)}}{5} z^5 \right) 
\end{equation} 
where we have used
Eq.~(\ref{gen.sol.1}) for the inhomogeneous term of Eq.~(\ref{diff2}). The
boundary conditions at $\mathcal{O}(Re)$ are 
\begin{equation}
\bar{\phi}^{(2)}_{0,-,1} (z=H) = 0 
\end{equation} 
\begin{equation} 
\partial_z
\bar{\phi}_{0,-,1}^{(2)} (z=H) = 0 
\end{equation} 
\begin{equation}
\bar{\phi}_{0,-,1}^{(1)} (z=-h) = 0 
\end{equation} 
\begin{equation} 
\partial_z
\bar{\phi}_{0,-,1}^{(1)} (z=-h) = 0 
\end{equation} 
\begin{equation}
\bar{\phi}^{(2)}_{0,-,1}(z=0) = \bar{\phi}^{(1)}_{0,-,1}(z=0) 
\end{equation}
\begin{equation} 
\partial_z u_{-,1}^{(2)}(z=0) \bar{z}_{s,0} + \partial_z
\bar{\phi}^{(2)}_{0,-,1}(z=0) = \partial_z u_{-,1}^{(1)}(z=0) \bar{z}_{s,0} +
\partial_z \bar{\phi}^{(1)}_{0,-,1}(z=0) 
\end{equation} 
\begin{equation}
G^{*(2)}(-\omega) \partial_z^2 \bar{\phi}_{0,-,1}^{(2)} (z=0) =
G^{*(1)}(-\omega) \partial_z^2 \bar{\phi}_{0,-,1}^{(1)} (z=0) 
\end{equation}
\begin{equation} 
G^{*(2)}(-\omega) \partial_z^3 \bar{\phi}_{0,-,1}^{(2)} (z=0) =
G^{*(1)}(-\omega) \partial_z^3 \bar{\phi}_{0,-,1}^{(1)} (z=0) 
\end{equation}
which can also be written as a 8$\times$8 matrix for the
coefficients of $\bar{\phi}_{0,-,1}^{(\alpha)}$. The solution is
\begin{eqnarray} 
E_0^{(1)} &=& E_0^{(2)} \nonumber \\ 
&=& i \frac{(n-1) n U
\zeta_{0} H^2}{60 (h+ nH)^2(n^2 H^4 + 4 n H^3 h + 6 n H^2 h^2 + 4 n h^3 H +
h^4)^2} \times \nonumber \\ 
&& \times (8 n^4 H^8 + 9 n^4 H^7 h + 68 n^3 H^7 h +
140 n^3 H^5 h^3 + 60 H^4 n^3 h^4 + 152 n^3 H^6 h^2 \nonumber \\ 
&& + 120 n^2 H^6
h^2 + 323 n^2 H^5 h^3 + 370 n^2 H^4 h^4 + 173 n^2 H^3 h^5 - 130 n H^3 h^5
\nonumber \\ 
&& - 82 h^7 n H -148 n H^2 h^6 -60 n H^4 h^4 -22 h^8 -21 h^7 H) 
\label{E0F1}
\end{eqnarray} 
\begin{eqnarray} 
F_0^{(1)} &=& - i \frac{n (n-1) H U
\zeta_{0}}{120 h^2 (h+ nH)^2(n^2 H^4 + 4 n H^3 h + 6 n H^2 h^2 + 4 n h^3 H +
h^4)^2} \times \nonumber \\ 
&& \times (10 n^5 H^{10} +80 H^7 n^4 h^3 +93 H^8 n^4
h^2 +78 n^4 H^9 h +160 H^4 n^3 h^6 \nonumber \\ 
&& +477 H^6 n^3 h^4 +128 n^3 H^8
h^2 +300 H^5 n^3 h^5 +305 H^7 n^3 h^3 +212 H^6 n^2 h^4 \nonumber \\ 
&& +588 H^3
n^2 h^7 +1003 H^4 n^2 h^6 +817 H^5 n^2 h^5 -535 h^8 n H^2 -497 H^3 n h^7
\nonumber \\ 
&& -328 h^9 n H -180 n h^6 H^4 -88 h^{10} -63 h^9 H) 
\label{F0F1}
\end{eqnarray} 
\begin{eqnarray} 
F_0^{(2)} &=& i \frac{(n-1) U \zeta_{0}}{120 h
(h+ nH)^2(n^2 H^4 + 4 n H^3 h + 6 n H^2 h^2 + 4 n h^3 H + h^4)^2} \times
\nonumber \\ 
&& \times (32 n^5 H^{10} +27 H^9 n^5 h +272 n^4 H^9 h +515 H^8 n^4
h^2 +463 H^7 n^4 h^3 \nonumber \\ 
&& +180 H^6 n^4 h^4 +480 n^3 H^8 h^2 +1068 H^7
n^3 h^3 +1123 H^6 n^3 h^4 +397 H^5 n^3 h^5 \nonumber \\ 
&& -28 H^4 n^3 h^6 -320
H^6 n^2 h^4 -540 H^5 n^2 h^5 -543 H^4 n^2 h^6 -175 H^3 n^2 h^7 \nonumber \\ 
&&
+8 h^8 n^2 H^2 -160 H^3 n h^7 -147 h^8 n H^2 -72 h^9 n H -20 h^{10})
\label{F0F2} 
\end{eqnarray} 
\begin{eqnarray} 
J_0^{(1)} &=& \frac{1}{n} J_0^{(2)}
\nonumber \\ 
&=& i \frac{(n-1) U \zeta_{0}}{30 h^2 (h+ nH)^2(n^2 H^4 + 4 n H^3 h
+ 6 n H^2 h^2 + 4 n h^3 H + h^4)^2} \times \nonumber \\ 
&& \times (4 n^5 H^{10}
+20 H^7 n^4 h^3 +34 n^4 H^9 h +27 H^8 n^4 h^2 -20 H^4 n^3 h^6 -16 H^6 n^3 h^4
\nonumber \\ 
&& +37 H^7 n^3 h^3 -81 H^5 n^3 h^5 +60 n^3 H^8 h^2 +22 H^3 n^2 h^7
-80 H^6 n^2 h^4 \nonumber \\ 
&& -96 H^5 n^2 h^5 -76 H^4 n^2 h^6 -48 h^8 n H^2
-41 h^9 n H -55 H^3 n h^7 -11 h^{10}) 
\label{J0F1} 
\end{eqnarray}
\begin{eqnarray} 
L_0^{(1)} &=& - i \frac{(n-1) U \zeta_{0}}{120 h^2 (h+
nH)^2(n^2 H^4 + 4 n H^3 h + 6 n H^2 h^2 + 4 n h^3 H + h^4)^2} \times \nonumber
\\ 
&& \times (9 n^5 H^9 +113 n^4 h H^8 +106 n^4 h^2 H^7 +60 n^4 h^3 H^6 +324 n^3
h^2 H^7 \nonumber \\ 
&& +522 n^3 h^3 H^6 +440 n^3 h^4 H^5 +64 n^3 h^5 H^4 +160
n^2 h^3 H^6 +176 n^2 h^4 H^5 \nonumber \\ 
&& +160 n^2 h^5 H^4 -42 n^2 h^6 H^3
+36 n^2 h^7 H^2 +140 n h^6 H^3 +134 n h^7 H^2 \nonumber \\ 
&& +127 n h^8 H +31
h^9) 
\label{L0F1} 
\end{eqnarray} 
\begin{eqnarray} 
L_0^{(2)} &=& i \frac{(n-1) n
U \zeta_{0}}{120 h^2 (h+ nH)^2(n^2 H^4 + 4 n H^3 h + 6 n H^2 h^2 + 4 n h^3 H +
h^4)^2} \times \nonumber \\ 
&& \times (n^5 H^9 -23 n^4 h H^8 +14 n^4 h^2 H^7 +20
n^4 h^3 H^6 -84 n^3 h^2 H^7 +78 n^3 h^3 H^6 \nonumber \\ 
&& +340 n^3 h^4 H^5
+416 n^3 h^5 H^4 +160 n^3 h^6 H^3 +304 n^2 h^4 H^5 +620 n^2 h^5 H^4 \nonumber \\
&& +642 n^2 h^6 H^3 +204 n^2 h^7 H^2 -60 n h^6 H^3 -14 n h^7 H^2 
-37 n h^8 H -21 h^9) 
\label{L0F2} 
\end{eqnarray}

\subsection{$\mathcal{O}(q_x)$}

At $\mathcal{O}(q_x)$, with $\sigma_0=\sigma_1=0$ the momentum conservation
equation becomes 
\begin{equation} 
\rho \partial_t \partial_z^2
\bar{\phi}_1^{(\alpha)} -\int_{-\infty}^t d t^\prime \; 
G^{(\alpha)}(t-t^\prime)
\partial_z^4 \bar{\phi}_1^{(\alpha)} = i \rho \bar{\phi}_0^{(\alpha)}
(\partial_z^2 u_b^{(\alpha)})
- i \rho u_b^{(\alpha)} \partial_z^2 \bar{\phi}_0^{(\alpha)} 
\end{equation}
Since we are interested in the time-independent non-periodic part of
$\bar{\phi}_1^{(\alpha)}$ in order to obtain the non-vanishing Floquet exponent
$\sigma_2$, $\partial_t \bar{\phi}_{1,\text{NP}} = 0$ and the momentum
conservation equation reduces to 
\begin{equation} \partial_z^4
\bar{\phi}_{1,\text{NP}}^{(\alpha)} = i \frac{\rho}{\eta^{(\alpha)}} \Big[
u_b^{(\alpha)} \partial_z^2 \bar{\phi}_0^{(\alpha)} - \bar{\phi}_0^{(\alpha)}
\partial_z^2 u_b^{(\alpha)} \Big]_\text{NP} 
\label{phi1.Oq} 
\end{equation} 
where the definition of the steady state viscosity is used 
\begin{equation}
\eta^{(\alpha)} = \int_0^{\infty} dt\; G^{(\alpha)}(t), 
\end{equation} 
and the non-periodic part of the RHS is given by 
\begin{equation} 
\Big[ u_b^{(\alpha)}
\partial_z^2 \bar{\phi}_0^{(\alpha)} - \bar{\phi}_0^{(\alpha)} \partial_z^2
u_b^{(\alpha)} \Big]_\text{NP} = \Big[ u_{+}^{(\alpha)} \partial_z^2
\bar{\phi}_{0,-}^{(\alpha)} + u_{-}^{(\alpha)} \partial_z^2
\bar{\phi}_{0,+}^{(\alpha)}
- \bar{\phi}_{0,-}^{(\alpha)} \partial_z^2 u_{+}^{(\alpha)} -
  \bar{\phi}_{0,+}^{(\alpha)} \partial_z^2 u_{-}^{(\alpha)} \Big].
\end{equation} 
Again, in the same way to calculate the solution at
$\mathcal{O}(1)$ in $q_x$ we take a solution expanding in small $Re$ as
Eq.~(\ref{eq:form.phi1NP}).  Thus, at first two orders in $Re$ it is the case
that
\begin{equation} 
\partial_z^4 \bar{\phi}_{1,\text{NP},0}^{(\alpha)}(z) = 0,
\label{diff3} 
\end{equation} 
\begin{equation} 
\partial_z^4
\bar{\phi}_{1,\text{NP},1}^{(\alpha)} = i\frac{n^{(\alpha)}}{\omega h^2} \Big[
u_{+,0}^{(\alpha)} \partial_z^2 \bar{\phi}_{0,-,0}^{(\alpha)} +
u_{-,0}^{(\alpha)} \partial_z^2. \bar{\phi}_{0,+,0}^{(\alpha)} \Big]
\label{diff4} 
\end{equation} 
The general solutions of Eqs.~(\ref{diff3}) and
(\ref{diff4}) are 
\begin{equation} 
\bar{\phi}_{1,\text{NP},0}^{(\alpha)}(z) =
A_1^{(\alpha)} + B_1^{(\alpha)} z + C_1^{(\alpha)} z^2 + D_1^{(\alpha)} z^3
\end{equation} 
\begin{equation} 
\bar{\phi}_{1,\text{NP},1}^{(\alpha)}(z) =
E_1^{(\alpha)} + F_1^{(\alpha)} z + J_1^{(\alpha)} z^2 + L_1^{(\alpha)} z^3 +
\frac{\Lambda_0^{(\alpha)}}{24}z^4 + \frac{\Lambda_1^{(\alpha)}}{120} z^5 +
\frac{\Lambda_2^{(\alpha)}}{360} z^6 
\end{equation} 
where 
\begin{equation}
\Lambda_1^{(\alpha)} = \frac{2in^{(\alpha)}}{\omega h^2} \Big(
[u_{-,0}^{(\alpha)}(0)]^{*} C_0^{(\alpha)} + \text{c.c.} \Big) 
\end{equation}
\begin{equation} 
\Lambda_2^{(\alpha)} = \frac{2in^{(\alpha)}}{\omega h^2} \Big(
[\partial_z u_{-,0}^{(\alpha)}(0)]^{*} C_0^{(\alpha)} + 3
[u_{-,0}^{(\alpha)}(0)]^{*} D_0^{(\alpha)} + \text{c.c} \Big) 
\end{equation}
\begin{equation} 
\Lambda_3^{(\alpha)} = \frac{6in^{(\alpha)}}{\omega h^2} \Big(
[\partial_z u_{-,0}^{(\alpha)}(0)]^{*} D_0^{(\alpha)} + \text{c.c.} \Big).
\end{equation} 
Similarly to the calculation at $\mathcal{O}(1)$ in $q_x$, the
boundary conditions for $\bar{\phi}_{1,\text{NP},0}^{(\alpha)}(z)$ are
\begin{equation} 
\bar{\phi}^{(2)}_{1,\text{NP},0} (z=H) = 0, 
\end{equation}
\begin{equation} 
\partial_z \bar{\phi}_{1,\text{NP},0}^{(2)} (z=H) = 0,
\end{equation} 
\begin{equation} 
\bar{\phi}_{1,\text{NP},0}^{(1)} (z=-h) = 0,
\end{equation} 
\begin{equation} 
\partial_z \bar{\phi}_{1,\text{NP},0}^{(1)} (z=-h) = 0, 
\end{equation} 
\begin{equation}
\bar{\phi}^{(2)}_{1,\text{NP},0}(z=0) = \bar{\phi}^{(1)}_{1,\text{NP},0}(z=0),
\end{equation} 
\begin{eqnarray} 
\partial_z \bar{\phi}^{(2)}_{1,\text{NP},0}(z=0)
- \partial_z \bar{\phi}^{(1)}_{1,\text{NP},0}(z=0) &=& \frac{1}{\omega} \Big[
  \partial_z u_{+,0}^{(1)} - \partial_z u_{+,0}^{(2)}(z=0) \Big]_{z=0} \Big[
z_{s,0} u_{-,0}^{(1)} + \bar{\phi}_{0,-,0}^{(1)} \Big]_{z=0} \nonumber \\ 
&& - \text{c.c.}, 
\end{eqnarray} 
\begin{equation} 
\partial_z^2
\bar{\phi}_{1,\text{NP},0}^{(2)} (z=0) = n \partial_z^2
\bar{\phi}_{1,\text{NP},0}^{(1)} (z=0), 
\end{equation} 
\begin{equation}
\partial_z^3 \bar{\phi}_{1,\text{NP},0}^{(2)} (z=0) = n \partial_z^3
\bar{\phi}_{1,\text{NP},0}^{(1)} (z=0), 
\end{equation} 
and for $\bar{\phi}_{1,\text{NP},1}^{(\alpha)}(z)$ 
\begin{equation}
\bar{\phi}^{(2)}_{1,\text{NP},1} (z=H) = 0, 
\end{equation} 
\begin{equation}
\partial_z \bar{\phi}_{1,\text{NP},1}^{(2)} (z=H) = 0, 
\end{equation}
\begin{equation} 
\bar{\phi}_{1,\text{NP},1}^{(1)} (z=-h) = 0, 
\end{equation}
\begin{equation} 
\partial_z \bar{\phi}_{1,\text{NP},1}^{(1)} (z=-h) = 0,
\end{equation} 
\begin{equation} \bar{\phi}^{(2)}_{1,\text{NP},1}(z=0) =
\bar{\phi}^{(1)}_{1,\text{NP},1}(z=0), 
\end{equation} 
\begin{eqnarray}
\partial_z \bar{\phi}^{(2)}_{1,\text{NP},1}(z=0) - \partial_z
\bar{\phi}^{(1)}_{1,\text{NP},1}(z=0) &=& \frac{1}{\omega} \Big[ \partial_z
u_{+,0}^{(1)} - \partial_z u_{+,0}^{(2)} \Big]_{z=0}\Big[ z_{s,0} u_{-,1}^{(1)}
+ \bar{\phi}_{0,-,1}^{(1)} \Big]_{z=0} \nonumber \\ 
&& +\frac{1}{\omega} \Big[
\partial_z u_{+,1}^{(1)} - \partial_z u_{+,1}^{(2)} \Big]_{z=0}\Big[ z_{s,0}
u_{-,0}^{(1)} + \bar{\phi}_{0,-,0}^{(1)} \Big]_{z=0} \nonumber \\ 
&&
- \text{c.c.}, 
\end{eqnarray} 
\begin{equation} 
\partial_z^2
  \bar{\phi}_{1,\text{NP},1}^{(2)} (z=0) = n \partial_z^2
\bar{\phi}_{1,\text{NP},1}^{(1)} (z=0), 
\end{equation} 
\begin{eqnarray}
\partial_z^3 \bar{\phi}_{1,\text{NP},1}^{(2)} (z=0) - n \partial_z^3
\bar{\phi}_{1,\text{NP},1}^{(1)} (z=0) &=& \frac{in}{\omega h^2}
\bar{\phi}_{0,-,0}^{(1)}(z=0) \Big( \partial_z u_{+,0}^{(1)} - \partial_z
u_{+,0}^{(2)} \Big)_{z=0} \nonumber \\ 
&& -\frac{in}{\omega h^2}
u_{+,0}^{(1)}(z=0) \Big( \partial_z \bar{\phi}_{0,-,0}^{(1)} - \partial_z
\bar{\phi}_{0,-,0}^{(2)} \Big)_{z=0} \nonumber \\ 
&&
- \text{c.c.}, 
\end{eqnarray} 
can also be separately written as a $8\times8$ matrix. The solutions are 
\begin{eqnarray} 
A_1^{(1)}
&=& A_1^{(2)} \nonumber \\ &=& i \frac{h^3 n^2 U^2 \zeta_{0} H^2 (H + h)
(g^{(2)} - g^{(1)})}{\omega (h+ nH)^2(n^2 H^4 + 4 n H^3 h + 6 n H^2 h^2 + 4 n
h^3 H + h^4)^3} \times \nonumber \\ 
&& \times ( 12 n H^3 h^5 + 14 n^2 h^6 H^2 +
8 h^7 n H + 16 n H^2 h^6 + 2 n^4 H^6 h^2 + 18 n^2 H^6 h^2 - 2 h^6 H^2 \nonumber
\\ 
&& + h^8 - 2 H^3 h^5 + n^4 H^8 + 2 n^4 H^7 h + 4 n^3 H^7 h + 2 n^2 H^7 h + 46
n^2 H^3 h^5 + 48 n^2 H^5 h^3 \nonumber \\ 
&& + 8 n^3 H^5 h^3 + 8 n^3 H^6 h^2 +
70 n^2 H^4 h^4 ) 
\label{A1F1} 
\end{eqnarray} 
\begin{eqnarray} 
B_1^{(1)} &=&
-\frac{(4 H^3 n +h^3 +3 n H^2 h)h}{n(	H^3 n +4 h^3 +3 h^2 H )H} B_1^{(2)}
\nonumber \\ &=& i \frac{n U^2 \zeta_{0} h^2 (4 H^3 n +h^3 +3 n H^2 h) (g^{(2)}
- g^{(1)})}{2\omega (h+ nH)^2(n^2 H^4 + 4 n H^3 h + 6 n H^2 h^2 + 4 n h^3 H +
h^4)^3} \times \nonumber \\ 
&& \times ( 8 H^5 n^3 h^3 +H^8 n^4 +70 H^4 n^2 h^4
+2 n^4 H^6 h^2 +12 H^3 n h^5 +48 H^5 n^2 h^3 \nonumber \\ 
&& +18 H^6 n^2 h^2 +2
h H^7 n^2 +8 H^6 n^3 h^2 +46 n^2 H^3 h^5 +4 H^7 n^3 h -2 h^5 H^3 \nonumber \\ 
&&
+h^8 +2 n^4 H^7 h +16 h^6 n H^2 -2 h^6 H^2 +14 h^6 n^2 H^2 +8 h^7 n H )
\label{B1F1} 
\end{eqnarray} 
\begin{eqnarray} 
C_1^{(1)} &=& \frac{1}{n} C_1^{(2)}
\nonumber \\ &=& i \frac{n U^2 \zeta_{0} h (H^3 n + h^3) (g^{(2)} -
g^{(1)})}{\omega (h+ nH)^2(n^2 H^4 + 4 n H^3 h + 6 n H^2 h^2 + 4 n h^3 H +
h^4)^3} \times \nonumber \\ 
&& \times ( 8 H^5 n^3 h^3 +H^8 n^4 +70 H^4 n^2 h^4
+2 n^4 H^6 h^2 +12 H^3 n h^5 +48 H^5 n^2 h^3 \nonumber \\ 
&& +18 H^6 n^2 h^2 +2
h H^7 n^2 +8 H^6 n^3 h^2 +46 n^2 H^3 h^5 +4 H^7 n^3 h -2 h^5 H^3 \nonumber \\ 
&&
+h^8 +2 n^4 H^7 h +16 h^6 n H^2 -2 h^6 H^2 +14 h^6 n^2 H^2 +8 h^7 n H )
\label{C1F1} 
\end{eqnarray} 
\begin{eqnarray} 
D_1^{(1)} &=& \frac{1}{n} D_1^{(2)}
\nonumber \\ &=& i \frac{n U^2 \zeta_{0} h (h^2 - n H^2) (g^{(2)} -
g^{(1)})}{2\omega (h+ nH)^2(n^2 H^4 + 4 n H^3 h + 6 n H^2 h^2 + 4 n h^3 H +
h^4)^3} \times \nonumber \\ 
&& \times ( 8 H^5 n^3 h^3 +H^8 n^4 +70 H^4 n^2 h^4
+2 n^4 H^6 h^2 +12 H^3 n h^5 \nonumber \\ && +48 H^5 n^2 h^3 +18 H^6 n^2 h^2 +2
h H^7 n^2 +8 H^6 n^3 h^2 +46 n^2 H^3 h^5 \nonumber \\ 
&& +4 H^7 n^3 h -2 h^5 H^3
+h^8 +2 n^4 H^7 h +16 h^6 n H^2 -2 h^6 H^2 \nonumber \\ 
&& +14 h^6 n^2 H^2 +8
h^7 n H ) 
\label{D1F1} 
\end{eqnarray} 
\begin{eqnarray} 
E_1^{(1)} &=& E_1^{(2)}
\nonumber \\ &=& i \frac{n (n-1) \zeta_{0} U^2 H^2}{120 \omega (h+ nH)^2(n^2 H^4
+ 4 n H^3 h + 6 n H^2 h^2 + 4 n h^3 H + h^4)^3} \times \nonumber \\ 
&&
\times(-224 h^8 n^3 H^4 + 766 h^{10} n H^2 + 304 n h^8 H^4 + 3884 h^7 n^2 H^5 +
43 h^{12} + 2268 h^9 n^2 H^3 \nonumber \\ 
&& 632 h^{10} n^2 H^2 -32 H^5 h^7 n^4
-16 H^8 h^4 n^5 + 760 h^9 H^3 n -12 h^3 H^9 n^5 + 9 h^4 H^8 n^4 \nonumber \\ 
&&
-64 h^5 n^4 H^7 + 332 h^{11} n H - 40 h^6 n^4 H^6 + 176 h^5 H^7 n^3 + 2040 h^6
H^6 n^2 -68 h^6 n^3 H^6 \nonumber \\ 
&& + 4027 h^8 n^2 H^4 - 288 h^7 n^3 H^5 + 4
n^6 h H^{11} + 22 n^5 h^2 H^{10} + 94 n^4 h^3 H^9 + 160 n^3 h^4 H^8 \nonumber 
\\
&& n^6 H^{12} + 38 h^{11} H + 8 n^5 h H^{11} + 24 n^4 h^2 H^{10} + 512 n^2 h^5
H^7) \label{E1F1} 
\end{eqnarray} 
\begin{eqnarray} 
F_1^{(1)} &=& i \frac{(n-1)
\zeta_{0} U^2}{120 h \omega (h+ nH)^2(n^2 H^4 + 4 n H^3 h + 6 n H^2 h^2 + 4 n
h^3 H + h^4)^3} \times \nonumber \\ 
&& \times(20 h^6 H^8 n^5 +452 H^6 h^8 n^4
+29 H^{12} h^2 n^6 +384 H^{10} h^4 n^4 \nonumber \\ 
&& +2372 H^5 h^9 n^2 +6 n^7
h H^{13} +242 H^3 h^{11} n +307 h^{12} n H^2 +5046 H^5 h^9 n^3 \nonumber \\ 
&&
+1488 h^{11} n^2 H^3 +1050 H^8 h^6 n^4 +103 H^{10} h^4 n^5 +1340 h^{10} H^4 n^3
+788 H^7 h^7 n^4 \nonumber \\ 
&& +928 H^6 h^8 n^2 +58 H^9 h^5 n^5 +48 n^5 h^2
H^{12} +166 H^{11} h^3 n^5 -20 n^6 h^3 H^{11} \nonumber \\ 
&& +16 n^6 h H^{13}
+154 h^{13} n H +8820 H^6 h^8 n^3 +292 h^{12} n^2 H^2 +240 h^9 H^5 n^4 \nonumber
\\ 
&& +2 n^7 H^{14} +2789 H^4 h^{10} n^2 -24 h^4 H^{10} n^6 +1120 H^9 h^5 n^3
-48 h^7 H^7 n^5 \nonumber \\ 
&& +20 h^{14} +828 H^9 h^5 n^4 +4432 H^8 h^6 n^3
+8536 H^7 h^7 n^3) 
\label{F1F1} 
\end{eqnarray}
\begin{eqnarray} 
F_1^{(2)} &=& i
\frac{n (n-1) \zeta_{0} U^2 H}{120 h \omega (h+ nH)^2(n^2 H^4 + 4 n H^3 h + 6 n
H^2 h^2 + 4 n h^3 H + h^4)^3} \times \nonumber \\ 
&& \times(-76 n^5 h^4 H^9 -848
n^4 h^6 H^7 -608 n^4 h^7 H^6 +2 n^6 h^2 H^{11} -160 n^4 h^8 H^5 \nonumber \\ 
&&
-664 n h^{12} H +10 n^6 h H^{12} -1337 n h^{10} H^3 -1020 n^3 h^8 H^5 -2400 n^3
h^7 H^6 \nonumber \\ 
&& -3556 n^2 h^7 H^6 -2782 n^3 h^6 H^7 +12 n^5 h H^{12}
-1264 n^2 h^{11} H^2 +16 n^3 h^9 H^4 \nonumber \\ 
&& -7100 n^2 h^8 H^5 -429 n^4
h^4 H^9 -768 n^2 h^6 H^7 -352 n^3 h^4 H^9 -7742 n^2 h^9 H^4 \nonumber \\ 
&& -822
n^4 h^5 H^8 -1540 n^3 h^5 H^8 +23 n^5 h^2 H^{11} -1406 n h^{11} H^2 -456 n h^9
H^4 \nonumber \\ 
&& -4434 n^2 h^{10} H^3 -60 n^4 h^3 H^{10} -86 h^{13} +32 n^4
h^2 H^{11} -34 n^5 h^3 H^{10} \nonumber \\ 
&& -32 n^5 h^5 H^8 +2 n^6 H^{13} -57
h^{12} H) 
\label{F1F2} 
\end{eqnarray} 
\begin{eqnarray} 
J_1^{(1)} &=& \frac{1}{n}
J_1^{(2)} \nonumber \\ &=& i \frac{(n-1) \zeta_{0} U^2}{120 \omega h^2 (h+
nH)^2(n^2 H^4 + 4 n H^3 h + 6 n H^2 h^2 + 4 n h^3 H + h^4)^3} \times \nonumber
\\ 
&& \times(-8 H^{12} h^2 n^6 +n^7 H^{14} +6008 H^7 h^7 n^3 -4 n^6 h^3 H^{11}
+2464 H^7 h^7 n^4 \nonumber \\ 
&& +1576 H^6 h^8 n^4 +2704 H^5 h^9 n^2 +480 h^9
H^5 n^4 +6093 H^6 h^8 n^3 +400 H^3 h^{11} n \nonumber \\ 
&& +3574 H^4 h^{10} n^2
+24 n^5 h^2 H^{12} +1224 H^9 h^5 n^4 +832 h^{10} H^4 n^3 +3080 H^8 h^6 n^3
\nonumber \\ 
&& +8 n^6 h H^{13} +524 h^{12} n H^2 +2148 h^{11} n^2 H^3 +332
h^{13} n H +336 H^{10} h^4 n^4 \nonumber \\ 
&& +56 H^{11} h^3 n^5 +1040 H^6 h^8
n^2 +800 H^9 h^5 n^3 +632 h^{12} n^2 H^2 +308 H^9 h^5 n^5 \nonumber \\ 
&& +3552
H^5 h^9 n^3 +160 h^6 H^8 n^5 +43 h^{14} +2391 H^8 h^6 n^4 +182 H^{10} h^4 n^5)
\label{J1F1} 
\end{eqnarray} 
\begin{eqnarray} 
L_1^{(1)} &=&
- i \frac{(n-1) \zeta_{0} U^2}{120 \omega h^2 (h+ nH)^2(n^2 H^4 + 4 n H^3 h + 6
  n H^2 h^2 + 4 n h^3 H + h^4)^3} \times \nonumber \\ 
&& \times(2 n^7 H^{13}
-252 n h^{10} H^3 -29 h^{13} -16 n^5 h^6 H^7 -1046 n^3 h^8 H^5 \nonumber \\ 
&&
-334 n h^{11} H^2 -90 n^5 h^3 H^{10} -684 n^4 h^4 H^9 -436 n^2 h^{11} H^2 -4 n^6
h^2 H^{11} \nonumber \\ 
&& -1104 n^3 h^5 H^8 -1896 n^3 h^7 H^6 +15 n^6 h H^{12}
-186 n^5 h^4 H^9 -900 n^4 h^7 H^6 \nonumber \\ 
&& -2060 n^3 h^6 H^7 -1296 n^2
h^10 H^3 -100 n^5 h^5 H^8 -1492 n^2 h^8 H^5 -226 n h^{12} H \nonumber \\ 
&& -156
n^3 h^9 H^4 -176 n^4 h^3 H^{10} -2059 n^2 h^9 H^4 -576 n^2 h^7 H^6 +8 n^5 h^2
H^{11} \nonumber \\ 
&& -320 n^3 h^4 H^9 -8 n^6 h^3 H^{10} -1367 n^4 h^5 H^8 -240
n^4 h^8 H^5 -1404 n^4 h^6 H^7) 
\label{L1F1} 
\end{eqnarray} 
\begin{eqnarray}
L_1^{(2)} &=&
- i \frac{n (n-1) \zeta_{0} U^2}{120 \omega h^2 (h+ nH)^2(n^2 H^4 + 4 n H^3 h +
  6 n H^2 h^2 + 4 n h^3 H + h^4)^3} \times \nonumber \\ 
&& \times(320 n^4 h^9
H^4 +2 n^7 H^{13} +800 n^3 h^{10} H^3 -152 n h^{10} H^3 -19 h^{13} \nonumber \\
&& +144 n^5 h^6 H^7 +4554 n^3 h^8 H^5 -174 n h^{11} H^2 +230 n^5 h^3 H^{10} +676
n^4 h^4 H^9 \nonumber \\ 
&& +64 n^2 h^{11} H^2 +16 n^6 h^2 H^{11} +496 n^3 h^5
H^8 +4184 n^3 h^7 H^6 +25 n^6 h H^{12} \nonumber \\ 
&& +334 n^5 h^4 H^9 +1940
n^4 h^7 H^6 +2060 n^3 h^6 H^7 +4 n^2 h^{10} H^3 +300 n^5 h^5 H^8 \nonumber \\ 
&&
-452 n^2 h^8 H^5 -106 n h^{12} H +2804 n^3 h^9 H^4 +144 n^4 h^3 H^{10} -389 n^2
h^9 H^4 \nonumber \\ 
&& -256 n^2 h^7 H^6 +108 n^5 h^2 H^{11} +12 n^6 h^3 H^{10}
+1583 n^4 h^5 H^8 +1040 n^4 h^8 H^5 \nonumber \\ 
&& +2236 n^4 h^6 H^7)
\label{L1F2} 
\end{eqnarray}

We now obtain the Floquet exponent Eq.~(\ref{eq:gen.sigma2}) using 
Eqs.~(\ref{A0F1}), (\ref{E0F1}), (\ref{A1F1}) and (\ref{E1F1}), resulting in
\begin{eqnarray}
\sigma_\text{el} &=& 
\frac{U^2 h^3 H^2 n (H+h) (n-1) (nH^2 - h^2) (nH^2 + h^2) (g^{(2)} - 
g^{(1)})}{\omega(h+nH)^2} \times
\nonumber
\\
&&
\times
\frac{n^2 H^4 + 2h n^2 H^3 + 2 H^2 n^2 h^2 + 2 n H^3 h + 4 n H^2 h^2 + 
4 h^3 n H + h^4}{(h^4 + 6 n H^ 2h^2 + n^2 H^4 + 4 h^3 n H + 4 n H^3 h)^3}
\label{eq:sigma2.elastic}
\end{eqnarray}
and
\begin{eqnarray}
\sigma_\text{vis} &=& 
\frac{U^2 H^2 n (n-1) (4h n^2 H^3 + n^2 H^4 + h^4 + 4 h^3 H + 
6 n H^2 h^2)}{120\omega (h + n H)^2(h^4 + 6 n H^ 2h^2 + n^2 H^4 + 4 h^3 n H + 
4 n H^3 h)^3} \times
\nonumber
\\
&&
\times
(n^4 H^8 + 4 n^3 H^7 h - 2 n^3 H^6 h^2 - 4 h^3 n^3 H^5 + 8 n^2 H^6 h^2 + 
8 n^2 H^5 h^3 
\nonumber
\\
&&
\hspace{0.5cm}
- 8 h^5 n^2 H^3 - 8 h^6 n^2 H^2 + 4 h^5 n H^3 + 2 h^6 n H^2 - 4 h^7 n H - h^8)
\label{eq:sigma2.viscous}
\end{eqnarray}

\section{Expansion for general Reynolds number}
\label{sec:appendixB}

Now we present a general framework to obtain a solution of the equations of
motion for an arbitrary Reynolds number, but still in the limit of small $q_x$.

\subsection{$\mathcal{O}(1)$}

For arbitrary Reynolds number we can find a general solution to
Eq.~(\ref{phi0.O1}) by taking Eq.~(\ref{phi0fnform}) for
$\bar{\phi}_{0}^{(\alpha)}$. For $e^{- i\omega t}$ mode the general solution to
Eq.~(\ref{phi0.O1}) is 
\begin{equation} 
\bar{\phi}_{0,-}^{(\alpha)}(z) =
\hat{A}_0^{(\alpha)} + \hat{B}_0^{(\alpha)} z + \hat{C}_0^{(\alpha)}
e^{k^{(\alpha)} z} + \hat{D}_0^{(\alpha)} e^{-k^{(\alpha)} z} 
\label{gen.phi0m}
\end{equation} 
where $k^{(\alpha)}$ is given by
Eq.~(\ref{base.flow.wavevector}). Again, the coefficients of
$\bar{\phi}_{0,-}^{(\alpha)}(z)$ are determined by applying the boundary
conditions, Eqs.~(\ref{2D.boundary.cond1}) - (\ref{2D.normal.stress}) which are
\begin{equation} 
\hat{A}_0^{(2)} + \hat{B}_0^{(2)} H + \hat{C}_0^{(2)}
e^{k^{(2)} H} + \hat{D}_0^{(2)} e^{-k^{(2)} H} = 0 
\end{equation}
\begin{equation} 
\hat{B}_0^{(2)} + k^{(2)} \hat{C}_0^{(2)} e^{k^{(2)} H} -
k^{(2)} \hat{D}_0^{(2)} e^{- k^{(2)} H} = 0 
\end{equation} 
\begin{equation}
\hat{A}_0^{(1)} - \hat{B}_0^{(1)} h + \hat{C}_0^{(1)} e^{-k^{(1)} h} +
\hat{D}_0^{(1)} e^{k^{(1)} h} = 0 
\end{equation} 
\begin{equation}
\hat{B}_0^{(1)} + k^{(1)} \hat{C}_0^{(1)} e^{-k^{(1)} h} - k^{(1)}
\hat{D}_0^{(1)} e^{ k^{(1)} h} = 0 
\end{equation} 
\begin{equation}
\hat{A}_0^{(2)} + \hat{C}_0^{(2)} + \hat{D}_0^{(2)} = \hat{A}_0^{(1)} +
\hat{C}_0^{(1)} + \hat{D}_0^{(1)} 
\end{equation} 
\begin{equation}
\frac{1}{2}\partial_z u_0^{(2)} (z=0) \zeta_0 + \hat{B}_0^{(2)} + k^{(2)}
\hat{C}_0^{(2)} - k^{(2)} \hat{D}_0^{(2)}  = \frac{1}{2}\partial_z u_0^{(1)}
(z=0) \zeta_0 + \hat{B}_0^{(1)} + k^{(1)} \hat{C}_0^{(1)} - k^{(1)}
\hat{D}_0^{(1)} 
\end{equation} 
\begin{equation} 
\hat{C}_0^{(2)} +
\hat{D}_0^{(2)} = \hat{C}_0^{(1)} + \hat{D}_0^{(1)} 
\end{equation}
\begin{equation} 
\hat{B}_0^{(2)} = \hat{B}_0^{(1)} 
\end{equation} 
Therefore we obtain $\bar{\phi}_0^{(\alpha)}$ by solving this 8$\times$8 
linear system.

\subsection{$\mathcal{O}(q_x)$}

At $\mathcal{O}(q_x)$, we have to solve Eq.~(\ref{phi1.Oq}) for $\bar{\phi}_{1,
\text{NP}}^{(\alpha)}$. If Eq.~(\ref{gen.phi0m}) is used in the RHS of
Eq.~(\ref{phi1.Oq}), we get 
\begin{eqnarray} 
\eta^{(\alpha)} \partial_z^4
\bar{\phi}_{1, \text{NP}}^{(\alpha)} &=& i\rho \Big[ (k^{(\alpha)})^2 -
(k^{(\alpha)*})^2 \Big] u_{+}^{(\alpha)} \Big[ \hat{C}_0^{(\alpha)}
e^{k^{(\alpha)} z} + \hat{D}_0^{(\alpha)} e^{- k^{(\alpha)} z} \Big] \nonumber
\\ 
&&
- i \rho (k^{(\alpha)*})^2 u_+^{(\alpha)} ( \hat{A}_0^{(\alpha)} +
  \hat{B}_0^{(\alpha)} z) - \text{c.c.} \label{2D.lin.OS.eqn.q.NP.1}
\end{eqnarray} 
The solution to the above differential equation has a homogeneous
solution $\bar{\phi}_{1,\text{NP},h}^{(\alpha)}$ and a particular solution
$\bar{\phi}_{1,\text{NP},p}^{(\alpha)}$. First, the particular solution can be
found by using the fact that 
\begin{equation} \partial_z^4 f(z) = (A + B
z)e^{\pm k z}, 
\end{equation} 
so that 
\begin{equation}
f_p(z)
= \Bigg( \frac{A}{k^4} \mp 4 \frac{B}{k^5} + \frac{B}{k^4} z \Bigg) e^{\pm k z},
\end{equation} 
and, similarly, 
\begin{equation} 
\partial_z^4 g(z) = C e^{\pm k z}, 
\end{equation} 
\begin{equation} 
g_p(z) = \frac{C}{k^4} e^{\pm k z}.
\end{equation} 
The resulting particular solution for fluid 1 is
\begin{eqnarray} 
\bar{\phi}^{(1)}_{1,\text{NP},p} &=& -\frac{i}{4}
\frac{\rho}{\eta^{(1)}} (k^{(1)*})^2 \frac{U}{K^{*}} \Bigg[
\frac{\hat{A}_0^{(1)}}{(k^{(1)*})^4} - 4 \frac{\hat{B}_0^{(1)}}{(k^{(1)*})^5} +
\frac{\hat{B}_0^{(1)}}{(k^{(1)*})^4} z \Bigg]  e^{k^{(1)*} (z+h)} \nonumber \\
&& +\frac{i}{4} \frac{\rho}{\eta^{(1)}} (k^{(1)*})^2 \frac{U}{K^{*}} \Bigg[
\frac{\hat{A}_0^{(1)}}{(k^{(1)*})^4} + 4 \frac{\hat{B}_0^{(1)}}{(k^{(1)*})^5} +
\frac{\hat{B}_0^{(1)}}{(k^{(1)*})^4} z \Bigg]  e^{- k^{(1)*} (z+h)} \nonumber \\
&& +\frac{i}{4} \frac{\rho}{\eta^{(1)}} \Big[(k^{(1)})^2 - (k^{(1)*})^2 \Big]
\frac{U}{K^{*}} e^{k^{(1)*} h} \nonumber \\ && \hspace{1cm} \times \Bigg[
\frac{\hat{C}_0^{(1)}}{(k^{(1)*} + k^{(1)})^4} e^{(k^{(1)*} + k^{(1)})z} +
\frac{\hat{D}_0^{(1)}}{(k^{(1)*} - k^{(1)})^4} e^{(k^{(1)*} - k^{(1)})z} \Bigg]
\nonumber \\ 
&& -\frac{i}{4} \frac{\rho}{\eta^{(1)}} \Big[(k^{(1)})^2 -
(k^{(1)*})^2 \Big] \frac{U}{K^{*}} e^{-k^{(1)*} h} \nonumber \\ 
&& \hspace{1cm}
\times \Bigg[ \frac{\hat{C}_0^{(1)}}{(k^{(1)*} - k^{(1)})^4} e^{-(k^{(1)*} -
k^{(1)})z} + \frac{\hat{D}_0^{(1)}}{(k^{(1)*} + k^{(1)})^4} e^{-(k^{(1)*} +
k^{(1)})z} \Bigg] \nonumber \\ 
&& - \text{c.c.} 
\end{eqnarray} 
and for fluid 2 
\begin{eqnarray}
\bar{\phi}^{(2)}_{1,\text{NP},p} &=& -\frac{i}{4} \frac{\rho}{\eta^{(2)}}
(k^{(2)*})^2 \frac{U}{K^{*}} \sinh(k^{(1)*} h) \Bigg[
\frac{\hat{A}_0^{(2)}}{(k^{(2)*})^4} - 4 \frac{\hat{B}_0^{(2)}}{(k^{(2)*})^5} +
\frac{\hat{B}_0^{(2)}}{(k^{(2)*})^4} z \Bigg] e^{k^{(2)*} z} \nonumber \\ 
&&
-\frac{i}{4} \frac{\rho}{\eta^{(2)}} (k^{(2)*})^2 \frac{U}{K^{*}} \sinh(k^{(1)*}
h) \Bigg[ \frac{\hat{A}_0^{(2)}}{(k^{(2)*})^4} + 4
\frac{\hat{B}_0^{(2)}}{(k^{(2)*})^5} + \frac{\hat{B}_0^{(2)}}{(k^{(2)*})^4} z
\Bigg] e^{-k^{(2)*} z} \nonumber \\ 
&& -\frac{i}{4} \frac{\rho}{\eta^{(2)}}
(k^{(2)*})^2 \frac{U}{K^{*}} \frac{G^{*(1)}(\omega) k^{(1)*}}{G^{*(2)}(\omega)
k^{(2)*}} \cosh(k^{(1)*} h) \Bigg[ \frac{\hat{A}_0^{(2)}}{(k^{(2)*})^4} - 4
\frac{\hat{B}_0^{(2)}}{(k^{(2)*})^5} + \frac{\hat{B}_0^{(2)}}{(k^{(2)*})^4} z
\Bigg] e^{k^{(2)*} z} \nonumber \\ 
&& +\frac{i}{4} \frac{\rho}{\eta^{(2)}}
(k^{(2)*})^2 \frac{U}{K^{*}} \frac{G^{*(1)}(\omega) k^{(1)*}}{G^{*(2)}(\omega)
k^{(2)*}} \cosh(k^{(1)*} h) \Bigg[ \frac{\hat{A}_0^{(2)}}{(k^{(2)*})^4} + 4
\frac{\hat{B}_0^{(2)}}{(k^{(2)*})^5} + \frac{\hat{B}_0^{(2)}}{(k^{(2)*})^4} z
\Bigg] e^{-k^{(2)*} z} \nonumber \\ 
&& +\frac{i}{4} \frac{\rho}{\eta^{(2)}}
\Big[(k^{(2)})^2 - (k^{(2)*})^2 \Big] \frac{U}{K^{*}} \sinh(k^{(1)*} h)
\nonumber \\ 
&& \hspace{0.5cm} \times \Bigg[ \hat{C}_0^{(2)} \Bigg(
\frac{e^{(k^{(2)*} + k^{(2)})z}}{(k^{(2)*} + k^{(2)})^4} + \frac{e^{(k^{(2)*} -
k^{(2)})z}}{(k^{(2)*} - k^{(2)})^4} \Bigg) + \hat{D}_0^{(1)} \Bigg(
\frac{e^{-(k^{(2)*} + k^{(2)})z}}{(k^{(2)*} - k^{(2)})^4} + \frac{e^{-(k^{(2)*}
- k^{(2)})z}}{(k^{(2)*} + k^{(2)})^4} \Bigg) \Bigg] \nonumber \\ 
&& +\frac{i}{4}
\frac{\rho}{\eta^{(2)}} \Big[(k^{(2)})^2 - (k^{(2)*})^2 \Big] \frac{U}{K^{*}}
\frac{G^{*(1)}(\omega) k^{(1)*}}{G^{*(2)}(\omega) k^{(2)*}} \cosh(k^{(1)*} h)
\nonumber \\ && \hspace{0.5cm} \times \Bigg[ \hat{C}_0^{(2)} \Bigg(
\frac{e^{(k^{(2)*} + k^{(2)})z}}{(k^{(2)*} + k^{(2)})^4} - \frac{e^{-(k^{(2)*} -
k^{(2)})z}}{(k^{(2)*} - k^{(2)})^4} \Bigg) + \hat{D}_0^{(1)} \Bigg(
\frac{e^{(k^{(2)*} + k^{(2)})z}}{(k^{(2)*} - k^{(2)})^4} - \frac{e^{-(k^{(2)*} -
k^{(2)})z}}{(k^{(2)*} + k^{(2)})^4} \Bigg) \Bigg] \nonumber \\ 
&& - \text{c.c.} 
\end{eqnarray} 
Note that $\bar{\phi}_{1,\text{NP},p}^{(\alpha)}(z)$ is purely imaginary.  
On the other hand, the homogeneous solution is 
\begin{equation}
\bar{\phi}^{(\alpha)}_{1,\text{NP},h} = \hat{A}_1^{(\alpha)} +
\hat{B}_1^{(\alpha)} z + \hat{C}_1^{(\alpha)} z^2 + \hat{D}_1^{(\alpha)} z^3
\end{equation} 
The eight coefficients of $\bar{\phi}^{(\alpha)}_{1,\text{NP},h}$
are determined by imposing the eight boundary conditions.  
\begin{equation}
\bar{\phi}_{1,\text{NP},h}^{(2)}(z=H) + \bar{\phi}_{1,\text{NP},p}^{(2)}(z=H) =
0 
\end{equation} 
\begin{equation} 
\partial_z
\bar{\phi}_{1,\text{NP},h}^{(2)}(z=H) + \partial_z
\bar{\phi}_{1,\text{NP},p}^{(2)}(z=H) = 0 
\end{equation} 
\begin{equation}
\bar{\phi}_{1,\text{NP},h}^{(1)}(z=-h) + \bar{\phi}_{1,\text{NP},p}^{(1)}(z=-h)
= 0 
\end{equation} 
\begin{equation} 
\partial_z
\bar{\phi}_{1,\text{NP},h}^{(1)}(z=-h) + \partial_z
\bar{\phi}_{1,\text{NP},p}^{(1)}(z=-h) = 0 
\end{equation} 
\begin{equation}
\bar{\phi}_{1,\text{NP},h}^{(2)}(z=0) + \bar{\phi}_{1,\text{NP},p}^{(2)}(z=0) =
\bar{\phi}_{1,\text{NP},h}^{(1)}(z=0) + \bar{\phi}_{1,\text{NP},p}^{(1)}(z=0)
\end{equation} 
\begin{equation} 
\begin{split} 
\Big[ \partial_z u_b^{(2)}(z=0) &
\bar{z}_{s,1} \Big]_\text{NP} + \partial_z \Big[
\bar{\phi}_{1,\text{NP},h}^{(2)} + \bar{\phi}_{1,\text{NP},p}^{(2)} \Big]_{z=0}
\\ &= \Big[ \partial_z u_b^{(1)}(z=0) \bar{z}_{s,1} \Big]_\text{NP} + \partial_z
\Big[ \bar{\phi}_{1,\text{NP},h}^{(1)} + \bar{\phi}_{1,\text{NP},p}^{(1)}
\Big]_{z=0} 
\end{split} 
\end{equation} 
\begin{equation} 
\eta^{(2)} \partial_z^2
\Big[  \bar{\phi}_{1,\text{NP},h}^{(2)} + \bar{\phi}_{1,\text{NP},p}^{(2)}
\Big]_{z=0} = \eta^{(1)} \partial_z^2 \Big[ \bar{\phi}_{1,\text{NP},h}^{(1)} +
\bar{\phi}_{1,\text{NP},p}^{(1)} \Big]_{z=0} 
\end{equation} 
\begin{equation}
\begin{split} &i \rho \Big[ u_b^{(2)} \partial_z \bar{\phi}_0^{(2)} -
\bar{\phi}_0^{(2)} \partial_z u_b^{(2)} \Big]_{\text{NP},z=0} - \eta^{(2)}
\partial_z^3 \Big[  \bar{\phi}_{1,\text{NP},h}^{(2)} +
\bar{\phi}_{1,\text{NP},p}^{(2)} \Big]_{z=0} \\ &= i \rho \Big[ u_b^{(1)}
\partial_z \bar{\phi}_0^{(1)} - \bar{\phi}_0^{(1)} \partial_z u_b^{(1)}
\Big]_{\text{NP},z=0} - \eta^{(1)} \partial_z^3 \Big[
\bar{\phi}_{1,\text{NP},h}^{(1)} + \bar{\phi}_{1,\text{NP},p}^{(1)} \Big]_{z=0}
\end{split} 
\end{equation} 
Again, the general solution can be found by solving
this 8$\times$8 matrix for the coefficients of
$\bar{\phi}_{1,\text{NP},h}^{(\alpha)}$

In principle, we can get $\sigma_2$ of Eq.~(\ref{sigma2}) from the derived
solutions $\bar{\phi}_{1,\text{NP},h}^{(\alpha)}$ and
$\bar{\phi}_{1,\text{NP},p}^{(\alpha)}$ 
\begin{equation} 
\sigma_2 = -i
\frac{q^2}{\zeta_0} \bigg\{ \Big[ u_b^{(1)}|_{z=0} \bar{z}_{s,1} \Big]_\text{NP}
+ \bar{\phi}_{1,\text{NP},h}^{(\alpha)}|_{z=0} +
\bar{\phi}_{1,\text{NP},p}^{(\alpha)}|_{z=0} \bigg\} 
\end{equation} 
where
\begin{equation} 
\Big[ u_b^{(\alpha)}|_{z=0} \bar{z}_{s,1}\Big]_\text{NP} =
\frac{1}{\omega} \Big\{ [u_{0,-}^{(1)}]^{*} \hat{A}_0^{(1)} - \text{c.c.}
\Big\}_{z=0} 
\end{equation} 
and 
\begin{equation}
\bar{\phi}_{1,\text{NP},h}^{(\alpha)}|_{z=0} = \hat{A}_1^{(1)} 
\end{equation}
\begin{eqnarray} 
\bar{\phi}_{1,\text{NP},p}^{(\alpha)}|_{z=0} &=& -\frac{i}{4}
\frac{\rho}{\eta^{(1)}} (k^{(1)*})^2 \frac{U}{K^{*}} e^{k^{(1)*} h} \Bigg[
\frac{\hat{A}_0^{(1)}}{(k^{(1)*})^4} - 4
\frac{\hat{B}_0^{(1)}}{(k^{(1)*})^5}\Bigg] \nonumber \\ 
&& +\frac{i}{4}
\frac{\rho}{\eta^{(1)}} (k^{(1)*})^2 \frac{U}{K^{*}} e^{-k^{(1)*} h} \Bigg[
\frac{\hat{A}_0^{(1)}}{(k^{(1)*})^4} + 4
\frac{\hat{B}_0^{(1)}}{(k^{(1)*})^5}\Bigg] \nonumber \\ 
&& +\frac{i}{4}
\frac{\rho}{\eta^{(1)}} \Big[(k^{(1)})^2 - (k^{(1)*})^2 \Big] \frac{U}{K^{*}}
e^{k^{(1)*} h} \Bigg[ \frac{\hat{C}_0^{(1)}}{(k^{(1)*} + k^{(1)})^4} +
\frac{\hat{D}_0^{(1)}}{(k^{(1)*} - k^{(1)})^4}\Bigg] \nonumber \\ 
&&
-\frac{i}{4} \frac{\rho}{\eta^{(1)}} \Big[(k^{(1)})^2 - (k^{(1)*})^2 \Big]
\frac{U}{K^{*}} e^{-k^{(1)*} h} \Bigg[ \frac{\hat{C}_0^{(1)}}{(k^{(1)*} -
k^{(1)})^4} + \frac{\hat{D}_0^{(1)}}{(k^{(1)*} + k^{(1)})^4}\Bigg] \nonumber \\
&& - \text{c.c.} 
\end{eqnarray} 
In general, $\sigma_2$ is too a complicated
function of fluid parameters to present explicitly, but we can readily
evaluate it numerically. However, one should be careful in the numerical 
evaluation as we have observed numerical instabilities in certain limits. For 
example, in the limit of small Reynolds number $\Re[k^{(\alpha)}]$ becomes 
too small, and causes singular behavior, leaving the numerical result 
unreliable. In this regard, we
decided to work directly on the limit of small Reynolds number as done in
Appendix~\ref{sec:appendixA}

\end{document}